\documentclass[twocolumn]{openjournal}
\usepackage{openjournal-submitted}
\usepackage{graphicx} 
\usepackage{multirow}

\usepackage[colorlinks,linkcolor=blue,citecolor=blue,urlcolor=blue ]{hyperref}
\usepackage[utf8]{inputenc}
\usepackage{float}
\usepackage{xcolor}
\usepackage{ulem}
\usepackage[T1]{fontenc}
\usepackage{amsmath}
\usepackage{xspace}
\usepackage{array,booktabs} 
\newcolumntype{P}[1]{>{\raggedright\arraybackslash}p{#1}} 

\newcommand{\msolar}{$\rm{M_{\odot}~}$}
\newcommand{\msolarc}{$\rm{M_{\odot}}$}
\newcommand{\msolaryr}{$\rm{M_{\odot} \ yr^{-1}}$}
\newcommand{\msolaryrc}{$\rm{M_{\odot} \ yr^{-1}}$}

\newcommand{\molH}{H$_{2}~$}

\newcommand{\rarepeak} {\textsc{Rarepeak}\xspace}
\newcommand{\NormalOne} {\textsc{Normal1}\xspace}
\newcommand{\NormalTwo} {\textsc{Normal2}\xspace}

\defcitealias{Prole2023}{LP23}
\defcitealias{Schauer2021}{AS21}

\begin{document}



\title{Black Hole Feedback, Galaxy Quenching and outflows at Cosmic Dawn: Analysis of the SEEDZ simulations\vspace{-1cm}}
\author{Lewis R. Prole$^{* \ 1}$}
\author{John A. Regan$^{1}$}
\author{Daxal Mehta$^{1}$}
\author{R\"udiger Pakmor$^{2}$}
\author{Sophie Koudmani$^{3,7}$}
\author{Martin A. Bourne$^{3,7,11}$}
\author{Simon C.~O. Glover$^{4}$}
\author{John H. Wise$^{5}$}
\author{Ralf S. Klessen$^{4,9}$}
\author{Michael Tremmel$^{6}$}
\author{Debora Sijacki$^{3,7}$}
\author{Ricarda S. Beckmann$^{10}$}
\author{Martin G. Haehnelt$^{3,7}$}
\author{John Brennan$^{1}$}
\author{Pelle van de Bor$^{1}$}
\author{Paul C. Clark$^{8}$}
\email{$^*$email: lewis.prole@mu.ie}
\affiliation{\\ \ \ \ \ $^{1}$ Centre for Astrophysics and Space Sciences Maynooth, Department of Physics, Maynooth University, Maynooth, Ireland.}
\affiliation{$^{2}$ Max-Planck-Institut f\"ur Astrophysik, Karl-Schwarzschild-Straße 1, D-85748, Garching, Germany}
\affiliation{$^{3}$ Institute of Astronomy, University of Cambridge, Madingley Road, Cambridge CB3 0HA, UK}
\affiliation {$^{4}$ Universit\"{a}t Heidelberg, Zentrum f\"{u}r Astronomie, Institut f\"{u}r Theoretische Astrophysik, Albert-Ueberle-Stra{\ss}e 2, 69120 Heidelberg, Germany.}
\affiliation {$^{5}$ Center for Relativistic Astrophysics, School of Physics, Georgia Institute of Technology, 837 State Street, Atlanta, GA 30332, USA}
\affiliation {$^{6}$ School of Physics, University College Cork, College Road, Cork T12 K8AF, Ireland}
\affiliation {$^{7}$ Kavli Institute of Cosmology, Cambridge, University of Cambridge, Madingley Road, Cambridge CB3 0HA, UK}
\affiliation {$^{8}$ Cardiff University School of Physics and Astronomy}
\affiliation{$^{9}$ Universit\"{a}t Heidelberg, Interdisziplin\"{a}res Zentrum f\"{u}r Wissenschaftliches Rechnen, Im Neuenheimer Feld 225, 69120 Heidelberg, Germany} 
\affiliation{$^{10}$ Institute for Astronomy, University of Edinburgh, Royal Observatory, Edinburgh EH9 3HJ, UK}
\affiliation{$^{11}$ Centre for Astrophysics Research, Department of Physics, Astronomy and Mathematics, University of Hertfordshire, College Lane, Hatfield, AL10 9AB, UK}

\begin{abstract}
\noindent 
Here we analyse the growth and feedback effects of massive black holes (MBHs) in the \texttt{SEEDZ} simulations. The most massive black holes grow to masses of $\sim10^{6}$ M$_\odot$ by $z=12.5$ during short bursts of super-Eddington accretion, sustained over periods of 5-30 Myr. We find that the determining factor that cuts off this initial growth is feedback from the MBH itself, rather than nearby supernovae or exhausting the available gas reservoir. Our simulations show that for the most actively accreting MBHs, feedback completely evacuates the gas from the host halo and ejects it into the inter-galactic medium. Despite implementing a relatively weak feedback model, the energy injected into the gas surrounding the MBH exceeds the binding energy of the halo. These results either indicate that MBH feedback in the early ($\Lambda$CDM) Universe is much weaker than previously assumed, or that at least some of the high redshift galaxies we currently observe with JWST formed via a two-step process, whereby a MBH  initially quenches its host galaxy and later reconstitutes its baryonic reservoir, either through mergers with gas rich galaxies or from accretion from the cosmic web. Moreover, the maximum black hole masses that emerge in \texttt{SEEDZ} are effectively set by a combination of MBH feedback modelling and the binding potential of the host halo. Unless feedback is extremely ineffective at early times (for example if growth is merger dominated rather than accretion dominated, or if feedback is contained within a small region around the MBH), our results suggest that the maximum mass of black holes at redshifts before 12.5 should not significantly exceed $10^6$ \msolarc. 
\end{abstract}

\keywords{}

\maketitle
\section{Introduction} 
\label{sec:intro}
\vspace{0.2cm}
\noindent The discovery by JWST of a substantial population of massive black holes (MBHs) of mass $>10^6$ M$_\odot$, existing within the first billion years of the Universe ($z \gtrsim 6$) \citep{Larson2023, Maiolino2024, Juodzbalis2024, Kovacs2024, Bogdan2024, Taylor2025} has challenged our understanding of compact object formation in the early Universe. Moreover, it is unclear how these black holes (BHs) grew from their initial seed masses by orders of magnitude within the allotted time frame. Current observations of MBHs now extend out to $z\sim 12.3$ \citep{Ortiz2025} and galaxy observations out to $z\sim 14.4$ \citep{Carniani2024,Naidu2026}. 
At these extreme redshifts, our understanding of galaxies, their small-scale properties, and their embedded black holes has previously relied heavily on theoretical models. However, the growing body of JWST observations now allows us to integrate theory and observational data from the early Universe. The \texttt{SEEDZ} simulation suite was designed to take advantage of this observational data, with the goal of understanding early galaxy and MBH evolution.

Did the first MBHs form from the remnants of the first Population III stars (so-called light seed black holes) and somehow sustain  Eddington growth rates throughout their lives? Or did short bursts of super-Eddington growth quickly amplify their masses \citep[e.g.][]{Madau2001, Milosavljevic2009, Alvarez2009, Lupi2016,Smith2018,Shi2023, Shi2024a, Gordon2024, Gordon2025, Zana2025, Mehta2024}? Was the required mass growth reduced by forming with higher initial masses (heavy seeds) via a number of connected formation pathways \citep{Regan2024a}? Heavy seed MBHs could be born from runaway collisions within a dense stellar cluster \citep{Begelman1978,PortegiesZwart2004,Reinoso2018, Reinoso2023,Rantala2025} or within a BH cluster \citep{Fragione2018}. Alternatively, heavy seeds may be born from the end points of super-massive stars (SMSs) or perhaps quasi-stars (QSs)  \citep{Loeb1994, Lodato2004, Begelman2006,Wise2008, Begelman2008,Regan2009,  Mayer2010, Agarwal2012, Agarwal2014, Habouzit2016, Inayoshi2015a, Ardaneh2018, Luo2018,  Regan2018, Visbal2014, Visbal2014a, Latif2013b, Regan2014,  Becerra2015,Regan2017, Wise2019, Regan2020, latif2021,  Chiaki2023, Prole2024a, Prole2024b, Cenci2025}. In this case, rapid accretion onto a protostar in excess of $\sim 10^{-2}$ \msolaryr{} is required to drive the formation of a bloated supermassive star with zero age main sequence masses between $10^3$ and $10^5$ \msolarc \citep[e.g.][]{Umeda2016,Woods2017,Haemmerle2018,Haemmerle2021,Nandal2023}. The eventual collapse of these objects after approximately 2 Myr is thought to result in the direct formation of a heavy seed MBH. Finally, heavy seeds have also been conjectured to form through the collisions of very massive halos at z $\lesssim$ 10 through the merger of gas rich and metal enriched halos of mass $\sim 10^{12}$ \msolar \citep{Mayer2010, Mayer2018, Mayer2024}. In this case, the MBH that directly forms can have an initial (seed) mass in excess of $10^{8}$ \msolarc. \\
\indent More exotically, it is conceivable that the seeds for the SMBH population formed earlier in cosmic history as primordial black holes, giving them more time to grow into SMBHs \citep{Kashlinsky2021, liu2022, Dayal2024, Prole2025, Ziparo2025, Zhang2025} with potential observable effects in the very first galaxies \citep[e.g.][]{Maiolino2025}

Irrespective of how these high redshift SMBHs formed, their current observed relationships with respect to their host galaxies show stark differences to established local Universe trends \citep{Pacucci2023, Pacucci2024, Dattathri2025}. Most notably, the local black hole-stellar mass (M$_{\rm BH}$-M$_{\rm star}$) relationship follows M$_{\rm BH}$/M$_{\rm star} \simeq 10^{-4}$ \citep[e.g.][]{Reines2015}, while high redshift observations produce relatively over-massive black holes with M$_{\rm BH}$/M$_{\rm star} \gtrsim 10^{-2}$ \citep[e.g.][]{Pacucci2023,Matthee2025} -- an effect we observed in the first paper in this series \citep{Prole2026}. This suggests that the stellar component of galaxies builds up after the formation of a MBH seed, transitioning from the high-redshift trends to the local relation over time as the host galaxy matures. Alternatively, the observed over-massive black hole to stellar mass ratios have been suggested to arise due to observational selection biases and measurement uncertainties, allowing only the so-called `tip of the iceberg' to be observed \citep{Li2024}.

The first suite of \texttt{SEEDZ} simulations \citep{Prole2026} were designed to investigate black hole formation and growth within a $\Lambda$CDM context, modeling light and heavy seed black holes in fully 3D hydrodynamic cosmological simulations, performed with the moving mesh code \texttt{Arepo2} \citep{Springel2010, Pakmor2016,Pakmor2023}. The simulations utilize new models for Population III star formation, supernovae (SNe) explosions and the resulting formation of light seed black holes, metal enrichment and subsequent Population II star formation, heavy seed black hole formation, Eddington and super-Eddington accretion schemes as well as black hole feedback. A potentially important consequence of our investigations would be the use of our results (based on sophisticated seeding strategies and a multiphase ISM model) in larger scale cosmological simulations which employ, by necessity, rather coarse resolution. \\
\indent In \cite{Prole2026}, we presented 3 simulations suites, referred to as \rarepeak, \NormalOne and \NormalTwo. We showed that a number of heavy seeds were able to grow from their initial seed masses of $\sim10^4$ M$_\odot$ into the upper $10^5$ M$_\odot$ regime by $z=15$, giving a number density of massive objects ($> 10^5$ M$_\odot$) of a few times 10$^{-2}$ cMpc$^{-3}$ h$^{-1}$. The simulations have now advanced to $z=12.5$, and at least 1 MBH per simulation has exceeded 10$^6$ M$_\odot$. We now aim to find the physical mechanism that causes a growing SMBH to stop accreting, and hence dictates its final mass at $z = 12.5$.

The paper is structures as follows; \S \ref{sec:method} summarizes the numerical methods relevant to BH growth used in the \texttt{SEEDZ} simulations. In \S \ref{sec:results} we present our results for MBH growth, galaxy quenching and outflows. In \S \ref{sec:conclusions}, we discuss these results and give our conclusions. Appendix \ref{sec:appendix} provides further evidence for the conclusions made.

\vspace{0.15cm}
\section{Numerical Method}
\vspace{0.2cm}
\label{sec:method}
\noindent \texttt{SEEDZ} uses the \texttt{Arepo2} cosmological code to model the early evolution of structure in the Universe. While a detailed description of the \texttt{SEEDZ} setup is given in \cite{Prole2026}, we briefly review the setup here. The initial \texttt{SEEDZ} suite is comprised of three zoom-in regions each of side length 6 cMpc h$^{-1}$, within a parent box of side length 40 cMpc h$^{-1}$. The three distinct regions are named Normal1, Normal2 and Rarepeak. As the names suggest, both the Normal1 and Normal2 regions are centred on regions of mean cosmic density, while the Rarepeak region targets an overdensity. More details on the structure of each region can be found in \cite{Prole2026}. The resolution of the zoom-in regions of these initial \texttt{SEEDZ} run is as follows:  
 the high resolution dark matter particle mass within each zoom region is approximately $6.9 \times 10^4$ \msolarc h$^{-1}$. The minimum cell size of the mesh is limited to 62 comoving pc h$^{-1}$ ($\sim$ 6 pc physical at $z=15$) and the minimum softening length of the dark matter and gas are set to 3.1 comoving kpc h$^{-1}$ ($\sim 300$ pc physical at $z = 15$) and 0.15 comoving kpc h$^{-1}$ ($\sim 14$ pc physical at $z = 15$) respectively. At this resolution, we are able to identify halos with masses a few times $10^6$ \msolar and we can spatially resolve the inner parts of galaxies down to scales of approximately 0.2 comoving kpc h$^{-1}$ (a few tens of parsecs physical at $z = 15$).  \\
 \indent Jeans refinement ensures that gas cells collapsing under gravity are resolved by 4 Jeans lengths down to the maximum refinement. Within gravitationally unstable gas cells at the maximum refinement level, a sink particle implementation allows for the formation of PopIII stars, PopII star clusters as well as heavy seed MBHs. The corresponding mass is removed from the host gas cell to form a sink particle. Heavy seed MBHs form with masses between $5 \times 10^3$ \msolar up to $10^5$ \msolar, sampling masses from a simple power-law distribution with a -0.5 slope, reflecting that the likelihood that more massive heavy seeds will be intrinsically rarer. MBHs are formed when the mass inflow onto a single cell exceeds 1 \msolaryrc. No restriction on metallicty is placed on forming a MBH, which means that MBHs can form in metal enriched gas, reflecting the theory that MBH seeds may emerge from collisions within dense stellar environments \cite[e.g.][]{Mapelli2016, Rantala2026} or that SMSs \citep[e.g.][]{Nandal2026}  may form in metal enriched environments.\\
 \indent PopII stellar clusters form when the gas is Jeans unstable, metal enriched above $10^{-4}$ Z$_\odot$ but with mass inflow rates below the threshold for heavy seed formation. The minimum PopII stellar mass is set at 1000 M$_\odot$ and at our resolution this results in typical PopII cluster masses of $10^4 - 10^5$ M$_\odot$. \\
 \indent PopIII stars form when neither the conditions for heavy seeds or PopII stellar clusters are achieved. Additionally, the \molH fraction must exceed $5 \times 10^{-4}$. The stellar mass of the PopIII star is randomly sampled from a top-heavy M$^{-1.3}$ initial mass function (IMF) in the range of 1-300 M$_{\odot}$ \citep[][]{Wise2011, Chen2014a, Prole2023}. Finally, PopIII stars (depending on their mass) can give birth to light seed black holes, following a supernovae event. Both light and heavy seed BHs are able to accrete from their surrounding gas cells. \\
\indent In summary \texttt{SEEDZ} contains three "particle" types in addition to dark matter particles and gas cells - PopIII stars, PopII stellar clusters and BHs, where BHs are conceptually split between light seeds formed via the PopIII progenitor channel, and heavy seeds formed via the heavy seed formation channel. Each of these particles are controlled by the sink particle framework that enables the \texttt{SEEDZ} model. The details of this framework can be found in \cite{Prole2026}. \\
\indent This paper focuses on the effects of heavy seed MBH growth and feedback on their host galaxies, and investigates what mechanism is responsible for halting MBH growth. In the following subsections we will cover the aspects of the \texttt{SEEDZ} simulations relevant to BH growth (the focus of this study): accretion, feedback, dynamical friction and BH mergers. For details on the formation of stars, SNe explosions and subsequent light seed BH formation, we again refer to the initial \texttt{SEEDZ} paper \citep{Prole2026}.

\subsection{Black Hole Accretion}
\label{sec:acc}
\vspace{0.2cm}
\noindent Upon formation, BH particles are assigned an accretion radius $R_{\rm acc}$ set to 5 times the chosen minimum cell length, giving $R_{\rm acc}=$ 0.62 ckpc h$^{-1}$ (for example, this is $\sim$45 pc physical at $z=20$). We model accretion onto BHs as Bondi-Hoyle-Lyttleton \citep{Bondi1952} accretion, following the formulation set out in \cite{Krumholz2004} and \cite{Krumholz2006}. The Bondi radius is given by
\begin{equation}
        R_{\rm Bondi} = \frac{G M_{\rm BH}}{v_{\infty}^2+c_{\infty}^2},
        \label{eq:bondi_radius}
\end{equation}
where $v_{\infty}$ is the mass-weighted speed of the gas within the accretion radius relative to the BH and $c_{\infty}$ is the sound speed in the region. 
The accretion rate onto the BH is then calculated using the Bondi formula as defined in \cite{Krumholz2004}
\begin{equation}
    \dot{M}_{\rm Bondi} = 4 \pi \rho_{\infty} R_{\rm Bondi}^2 ((1.12 c_{\infty})^2 + v_{\infty}^2)^{1/2}
    \label{eq:bondi-hoyle-acc}
\end{equation}
where $\rho_{\infty}$ is the weighted density for each cell inside the accretion radius, computed as described below.

As the Bondi radius scales with the square of the black hole mass, we do not always resolve the Bondi radius.
For lower mass light seed black holes this becomes particularly challenging, hence we do not resolve the Bondi radius of these black holes in any of the simulations presented in this paper (see \citealt{Mehta2026} for a discussion on this point). In order to account for this, we follow \cite{Krumholz2004} and define a kernel radius which accounts for the resolution of the simulation relative to the Bondi radius. The kernel radius, r$_{\rm K}$, is computed as:
\begin{equation}
r_{\rm K} = \left\{ \begin{array}{lcr}
  \Delta x_{\rm min}/4 & &R_{\rm Bondi} <  \Delta x_{\rm min}/4\\
  R_{\rm Bondi}  & & \ \ \ \ \Delta x_{\rm min}/4 \le R_{\rm Bondi} \le R_{\rm acc}/2\\
  R_{\rm acc}/2 && R_{\rm Bondi} > R_{\rm acc}/2
\end{array} \right.
\end{equation}
where $\Delta x_{\rm min}$ is the global minimum cell radius, estimated from the cell volume $V$ as $(3V / 4 \pi)^{1/3}$. A Gaussian kernel radius is used to assign a weight to every cell within $R_{\rm acc}$ using
\begin{equation}
    W \propto \exp(-r^2/r_{\rm K}^2),
\label{eq:weights}
\end{equation}
where $r$ is the distance from the cell to the accreting BH. The weighted density $\rho_{\infty}$ is then calculated as:
\begin{equation}
     \rho_{\infty} = \bar{\rho}  W
\end{equation}
where $\bar{\rho}$ is the mass-weighted mean density within the accretion sphere. 

Alternative models to the BHL accretion prescription have been explored extensively in the literature. For example, \cite{Hopkins2011} proposed a torque-limited model in which BH growth is regulated by gravitational instabilities on galactic scales, an approach subsequently implemented in cosmological simulations \citep{AnglesAlcazar2015, AnglesAlcazar2017}. Other works have incorporated angular momentum support and multiphase gas structure into modified accretion prescriptions \citep[e.g.][]{Booth2009, RosasGuevara2015, Regan2018a}. \\
\indent These models relax some of the key assumptions of the BHL formalism, in particular spherical symmetry and negligible angular momentum, and can lead to systematically different accretion rates depending on the dynamical state of the gas. In the dense, rapidly evolving, and strongly non-spherical environments studied here, angular momentum support and turbulent structure may act to suppress accretion relative to the BHL estimate, while large-scale gravitational torques or cold gas inflows may enhance it. We therefore caution that the inferred black hole growth rates may be sensitive to the adopted sub-grid prescription, and a systematic comparison of alternative accretion models will be the subject of future work, but is outside the scope of the present study.

\subsubsection{Vorticity Adjustment}
\vspace{0.2cm}
\noindent Following \cite{Krumholz2006} and \cite{Mehta2024}, we further adjust the accretion rate based on the vorticity $\omega$ of the surrounding gas. Low angular momentum gas is preferentially much easier to accrete and accounting for the vorticity of the gas addresses over-accretion of high angular momentum gas. The vorticity of the gas is given by
\begin{equation}
    \omega = | \nabla \times \vec{v} |
\end{equation}
of which we take the mass weighted average within the accretion sphere. We then calculate the dimensionless vorticity $\omega_*$ as
\begin{equation}
\omega_* =  \omega  \frac{R_{\rm Bondi}}{c_{\infty}}.
\end{equation}
and introduce a damping factor $f(\omega)$ defined as
\begin{equation}
    f_{w} = \frac{1}{1 + \omega_*^{0.9}}.
\end{equation}
The accretion rate in a turbulent medium can then be calculated according to
\begin{equation}
    \dot{M}_{\omega} = 4  \pi  \rho_{\infty}  R_{\rm Bondi}^2  c_{\infty}  (0.34  f_{\omega_*}).
\end{equation}
The total accretion rate onto the BH particle is
\begin{equation}
    \dot{M} = (\dot{M}_{\rm Bondi}^{-2} + \dot{M}_{\omega}^{-2})^{-0.5}.
\end{equation}

For a given time step, $t_h$, the mass of the BH particle increases by $M_{\rm acc} = t_h \dot{M}$, which is removed from cells within $R_{\rm acc}$ using the weighting scheme calculated in Equation \ref{eq:weights}, adjusting the velocities of both the gas cells and the BH to conserve linear momentum. The BH position is then shifted to the center of mass of a system comprised of the BH and the accreted mass contributions from each cell.

\subsection{Black Hole Accretion Feedback}
\label{sec:feedback_method}
\vspace{0.2cm}
\noindent To account for BH accretion feedback, we inject thermal energy into gas cells within the accretion region isotropically, based on the accretion rate onto a BH. The total mass accreted during a single timestep is given by $M_{\rm acc}$. This produces an amount of thermal energy
\begin{equation}
\label{eq:feedback}
    E_{\rm th, acc} = \epsilon f_{\rm c}  M_{\rm acc}c^2,
\end{equation}
where $f_{\rm c}$ is the thermal coupling factor, which we set to 0.05 \citep[see e.g.][]{Booth2009}, and $\epsilon$ is the radiative efficiency. The calculation of $\epsilon$ changes based on whether the BH is accreting at sub- or super-Eddington rates. For sub-Eddington accretion rates, we calculate $\epsilon$ as
\begin{equation}
    \epsilon = 1 - \sqrt{1-\frac{2}{3 R_{\rm ISCO}}},
\label{eq:epsilon}
\end{equation}
where $R_{\rm ISCO}$ is the innermost stable orbit. Assuming the BH's spin is prograde with respect to the accretion disc, $R_{\rm ISCO}$ is given by
\begin{equation}
    R_{\rm ISCO} = 3 + r_2 - \sqrt{(3 - r_1) \times (3 + r_1 + 2 \times r_2)},
\end{equation}
in units of $GM/c^2$ and $r_1$ and $r_2$ are given by
\begin{align}
r_1 &= 1 + (1 - a^2)^{1/3}  (1 + a)^{1/3} + (1 - a)^{1/3}, \\
r_2 &= \sqrt{3  a^2 + r_1^2},
\end{align}
and $a$ is the dimensionless black hole spin, assumed to be 0.7. These assumptions lead to a radiative efficiency $\epsilon$ $\sim$ 0.1.

When the accretion rate exceeds the Eddington limit, we transition to a slim disc model following \cite{Madau2014}. In this regime, photon trapping becomes important, such that a significant fraction of the generated radiation is advected inward with the flow rather than escaping. This reduces the effective radiative efficiency (and hence $\epsilon$) relative to the sub-Eddington case, weakening the resulting radiative feedback.

We calculate the Eddington accretion limit as
\begin{equation}
    \dot{M}_{\rm Edd}=\frac{4 \pi G M_{\rm BH} m_p}{ \epsilon c \sigma_{\rm T}},
\end{equation}
where $G$ is the gravitational constant, $c$ is the speed of light, $\epsilon$ is the sub-Eddington value calculated using Equation \ref{eq:epsilon} and $\sigma_{\rm T} = 6.65 \times 10^{-25} \: {\rm cm^{2}}$ is the Thomson scattering cross-section for an electron. 

In the super-Eddington regime, we first calculate the Eddington luminosity as
\begin{equation}
    L_{\rm Edd}=\frac{4 \pi G M_{\rm BH} m_p \mu c}{\sigma_{\rm T}},
\end{equation}
where $\mu$ is the mean molecular weight, assumed to be 1.22. We then use the parametrization found in \citet{Madau2014} to calculate the super-Eddington luminosity
\begin{equation}
    L_{\rm SE}=L_{\rm Edd} A \left(  \frac{0.985}{r_{\rm Edd} + B} + \frac{0.015}{r_{\rm Edd} + C} \right),
\end{equation}
where $r_{\rm Edd}$ is the ratio of the accretion rate to the Eddington limit, and
\begin{equation}
\begin{aligned}
A &= (0.9663 - 0.9292  a)^{-0.5639},\\
B &= (4.627 - 4.445  a)^{-0.5524},\\
C &= (827.3 - 718.1  a)^{-0.7060},\\
\end{aligned}
\end{equation}
where again the BH spin $a$ is set to 0.7. The value of $\epsilon$ in the super-Eddington case then follows as
\begin{equation}
\epsilon  = L_{\rm SE} / (\dot{M}_{\rm BH} c^2).
\end{equation}
Implementing this parameterisation reduces the value of $\epsilon$ for super-Eddington cases, thereby modulating the effective feedback injection at high accretion rates. We calculate the super-Eddington feedback energy by using the updated value of $\epsilon$ in Equation \ref{eq:feedback}. For numerical stability, we limit the amount of energy that can be injected into a gas cell in any one timestep to be five times the current internal energy of the cell. We note that this may artificially enhance accretion, or allow super-Eddington accretion for longer periods, thereby overestimating the maximum BH masses. The upcoming suite of higher resolution \texttt{SEEDZ} simulations will naturally produce higher density gas surrounding BHs, and will therefore be less reliant on this artificial energy limit.

Real AGN feedback is expected to be anisotropic and include kinetic components (e.g. jets or winds). Our isotropic thermal energy prescription likely maximises the coupling of the injected energy to the surrounding gas, and represents an upper limit on the effective coupling efficiency of BH feedback.

\begin{figure*}
  \centering
  
  \includegraphics[width=0.7\linewidth]{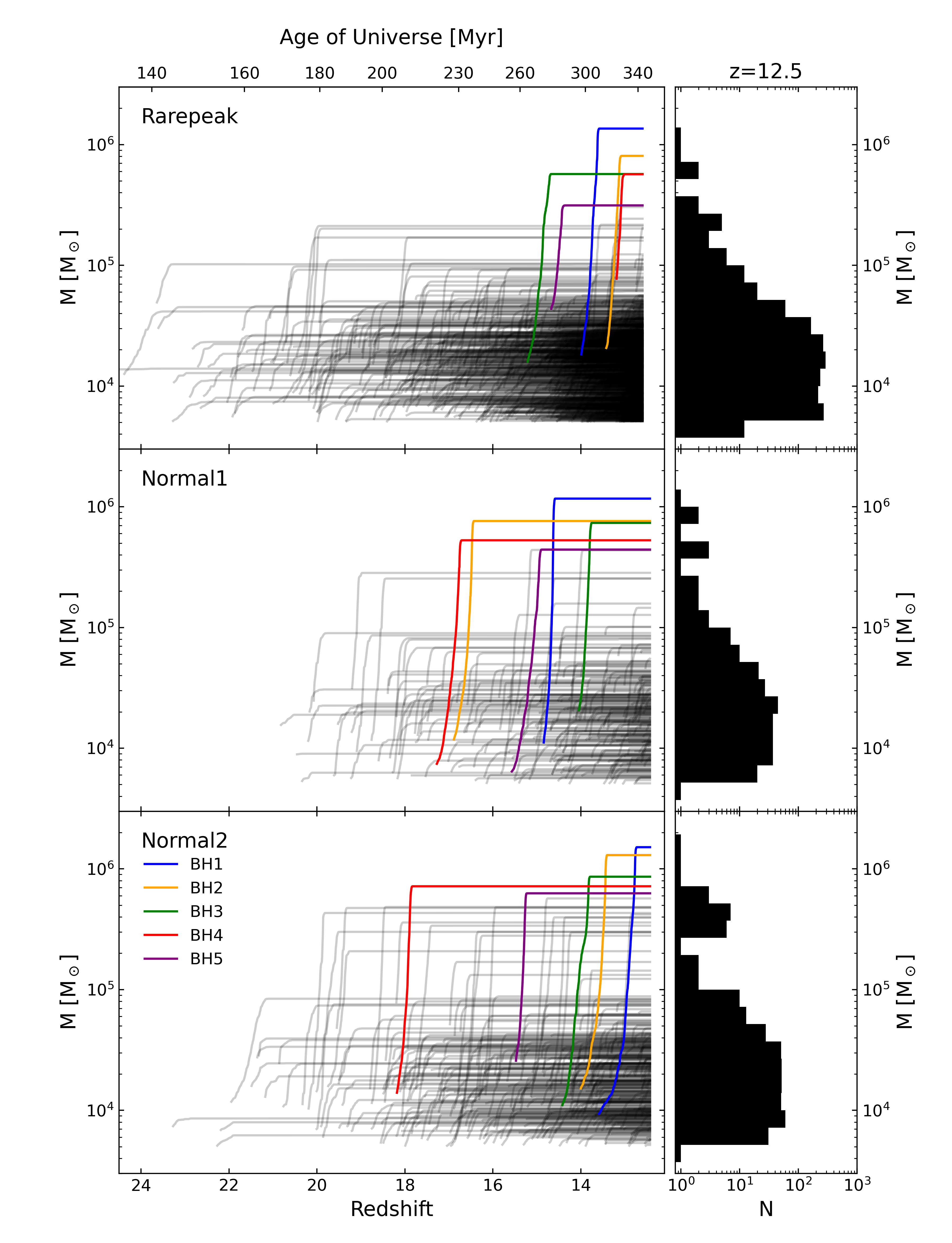}
 
\caption[]{\label{fig:growth} The growth of the most massive black holes across each of the three \texttt{SEEDZ} volumes (\rarepeak,\NormalOne, \NormalTwo). Each simulation has now reached $z = 12.5$, with the most massive MBH in each realization now above $10^6$ \msolarc. The majority of MBHs however lie within a factor of a few of their initial seed mass. We show the growth history of the top 5 most massive MBHs in each realisation as colored lines.\\}
\end{figure*}

\begin{figure}
  \centering
  \includegraphics[width=1\linewidth]{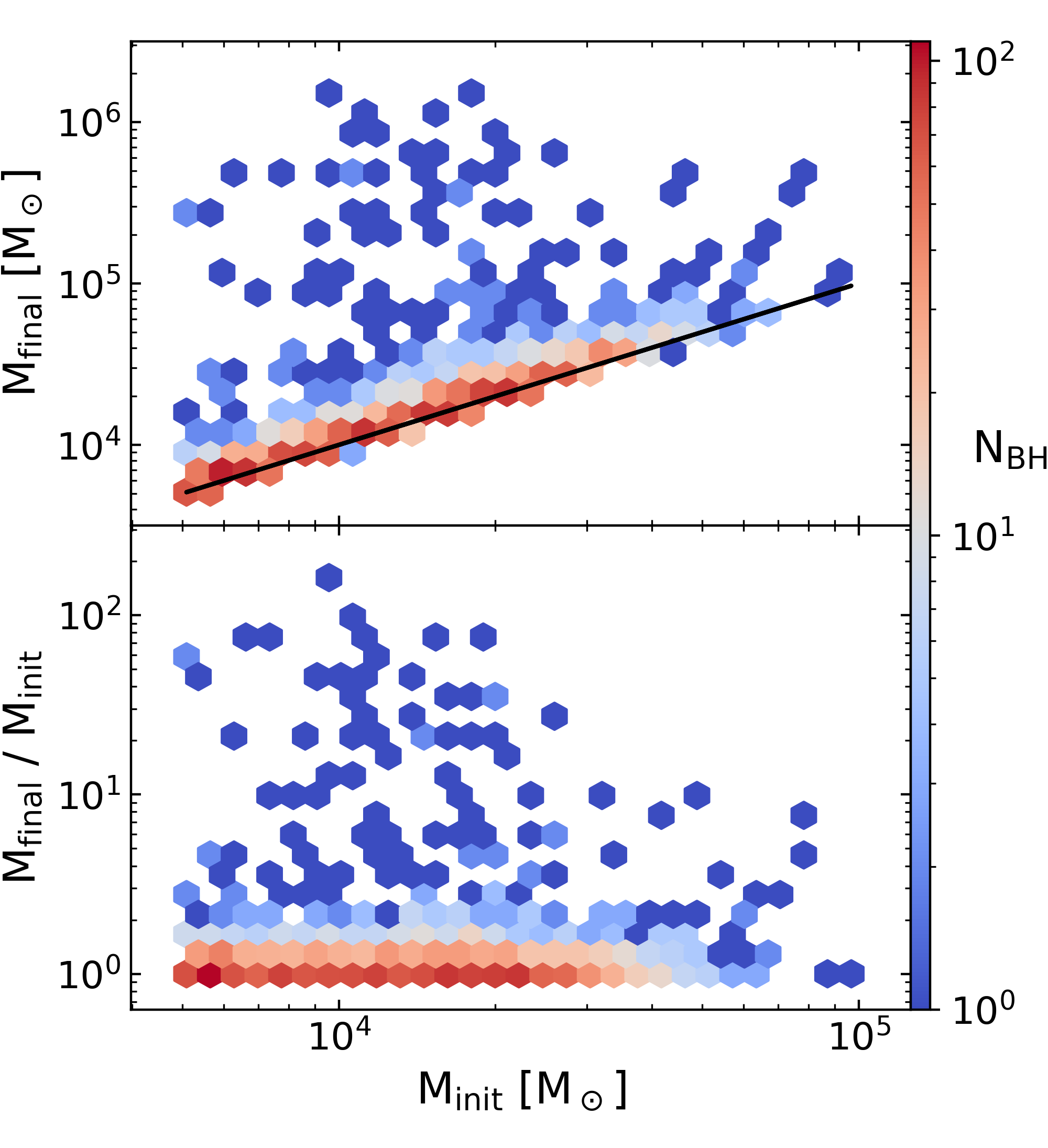}
\caption[]{\label{fig:initial_mass} Hexbin (2D histogram) showing the relation between BH initial and final ($z=12.5$) masses, for all heavy seeds (top). The black line shows zero growth i.e. M$_{\rm=init}$ = M$_{\rm final}$. We also express this relation as the ratio between final and initial mass (bottom). The colourbar shows the number of BHs within each hexbin.\\}
\end{figure}

\begin{figure*}
  \centering
  
  \includegraphics[width=1\linewidth]{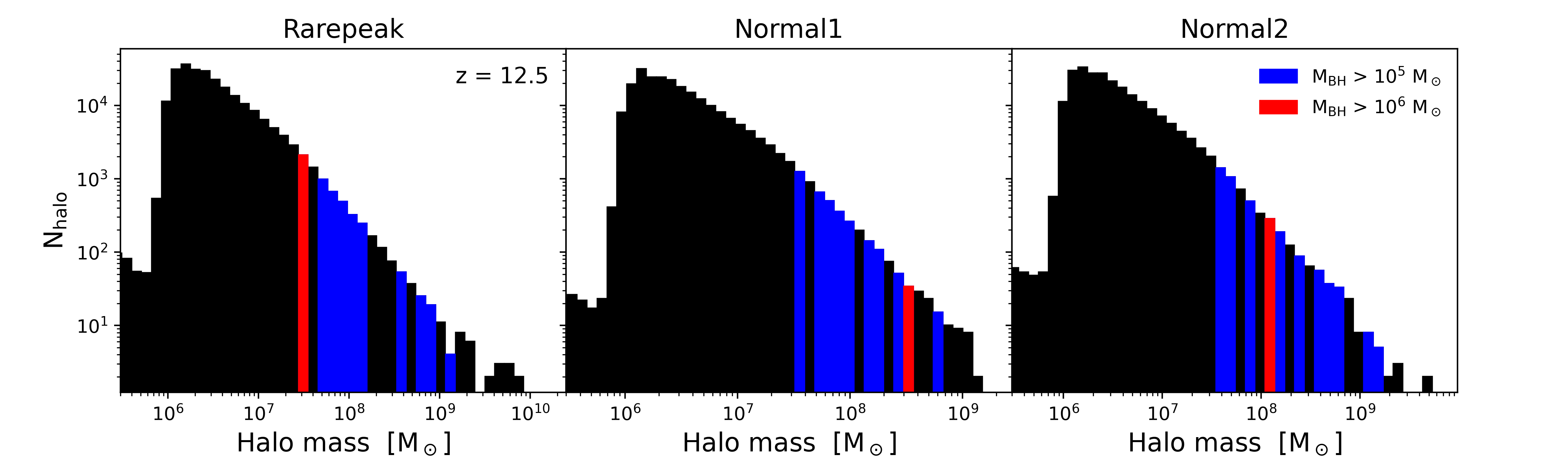}
 
\caption[]{\label{fig:hmf} Halo mass functions (HMFs) for the 3 simulation regions at $z=12.5$. We colour mass bins containing MBHs of mass greater than 10$^5$ and 10$^6$ M$_\odot$ in blue and red, respectively. Note that the heights of these coloured bins have not been altered from the original HMF, therefore the heights do not indicate number of halos containing MBHs.\\ \\}
\end{figure*}


\subsection{Dynamical Friction}
\vspace{0.2cm}
\noindent Dynamical friction is the drag force experienced by heavy particles moving through a background sea of lighter dark matter particles, stars, and gas, causing the orbits to decay towards the gravitational potential minima \citep{Chandrasekhar1943, Binney2011}. In cosmological simulations, large dark matter particle masses make it hard to resolve gravitational friction and also leads to unphysical gravitational heating, due to two-body encounters between particles. In order to avoid this numerical heating, gravitational interactions between particles are damped below a specified gravitational softening length \citep{Governato1994, Kazantzidis2005}. However, this has the side effect of reducing the dynamical friction acting on the heavy particles. To correct for this effect, we include an additional acceleration term acting on the BH particles, that accounts for gravitational interactions that are artificially suppressed by the softening. The volume over which interactions are suppressed is $V_{\rm s} = 4/3 \pi r_s^3$,
where $r_s=3.1$ ckpc/h is the softening length for BH particles. This softening remains constant and does not change with time or local properties around the BH particles. The softening length for BH particles is $\sim 20$ times larger than the minimum resolution of gas cells.

\indent We build our dynamical friction model based on \cite{Tremmel2015}, adding it to \texttt{Arepo2}.  We assume that the velocity dispersion of particles within the radius $r_s$ is isotropic and that the classical Chandrasekhar treatment of dynamical friction applies \citep{Chandrasekhar1943}. We further assume that only particles moving slower than the BH would influence its motion, giving us an acceleration of the form
\begin{equation}
    a_{\rm DF} = - 4 \pi G^2 M_{\rm BH} \rho(<v_{\rm BH})\ln{\Lambda} \frac{v_{\rm BH}}{|v_{\rm BH}^3|}
\end{equation}
where $\rho(<v_{\rm BH})$ is the density of background particles moving slower than than the BH and $v_{\rm BH}$ is the velocity of the BH. The Coulomb logarithm, $\ln{\Lambda}$,  depends on the maximum and minimum impact parameters, $b_{\rm max}$ and $b_{\rm min}$, as $\Lambda = b_{\rm max} /b_{\rm min}$. Since gravitational interactions beyond the softening length are well resolved, we take $b_{\rm max} = r_{\rm s}$ in order to avoid any over-counting of the effects of dynamical friction. For the minimum impact parameter, we take the radius of the influence for the BH, i.e.
\begin{equation}
    b_{\rm min} =  \frac{GM_{\rm BH}}{v_{\rm BH}^2}.
\end{equation}
For the purpose of this study, only dynamical friction from dark matter particles is taken into consideration when computing this correction. The dynamical friction from stars will be small at our current resolution and was neglected for this work, but will be added in future versions of the model. In addition, we only employ dynamical friction for BHs whose mass is larger than 5 times the dark matter particle mass ( $\sim 3.45 \times 10^5 h^{-1}$ \msolarc), as the dynamics can be hindered further by applying the correction when M$_{\rm BH} \sim $M$_{\rm DM}$. For BHs with masses comparable to individual dark matter particles, the resulting dynamical friction becomes noisy and dominated by stochastic fluctuations \citep{Tremmel2015}. The resulting acceleration is added to the BH acceleration to be integrated over the next time-step. As the mass of the BH increases, the radius of influence grows and can eventually become larger than $b_{\rm max}$. Once this happens, we assume that dynamical friction is fully resolved and no longer include this correction term.

\vspace{0.1cm}
\subsection{Stellar and Black Hole Mergers}
\label{sec:mergers}
\vspace{0.2cm}
\noindent We allow black hole particles to merge based on the treatment originally implemented in \cite{Prole2022}. Black hole particles are merged if:
   \begin{itemize}
      \item They lie within each other’s accretion radius of 0.62 ckpc h$^{-1}$. 
      \item They are moving towards each other.
      \item Their relative accelerations are $<$0.
      \item They are gravitationally bound to each other.
   \end{itemize}
As our BH particles carry no thermal data, the last criterion simply requires that their gravitational potential exceeds the kinetic energy of the system. When these criteria are met, the larger of the particles gains the mass and linear momentum of the smaller particle, and its position is shifted to the centre of mass of the system. Comparing the accretion radius of 0.62 ckpc h$^{-1}$ to the sink particle gravitational softening length of 3.1 ckpc h$^{-1}$ means that mergers will likely be somewhat suppressed within this initial suite of `low resolution' calibration runs. This initial set of simulations is designed to test the limitations of our modelling and hence we dub them the "Calibration" set. A later set of simulations will attempt to build on this initial suite.

\begin{figure*}
  \centering
  
  \includegraphics[width=\linewidth]{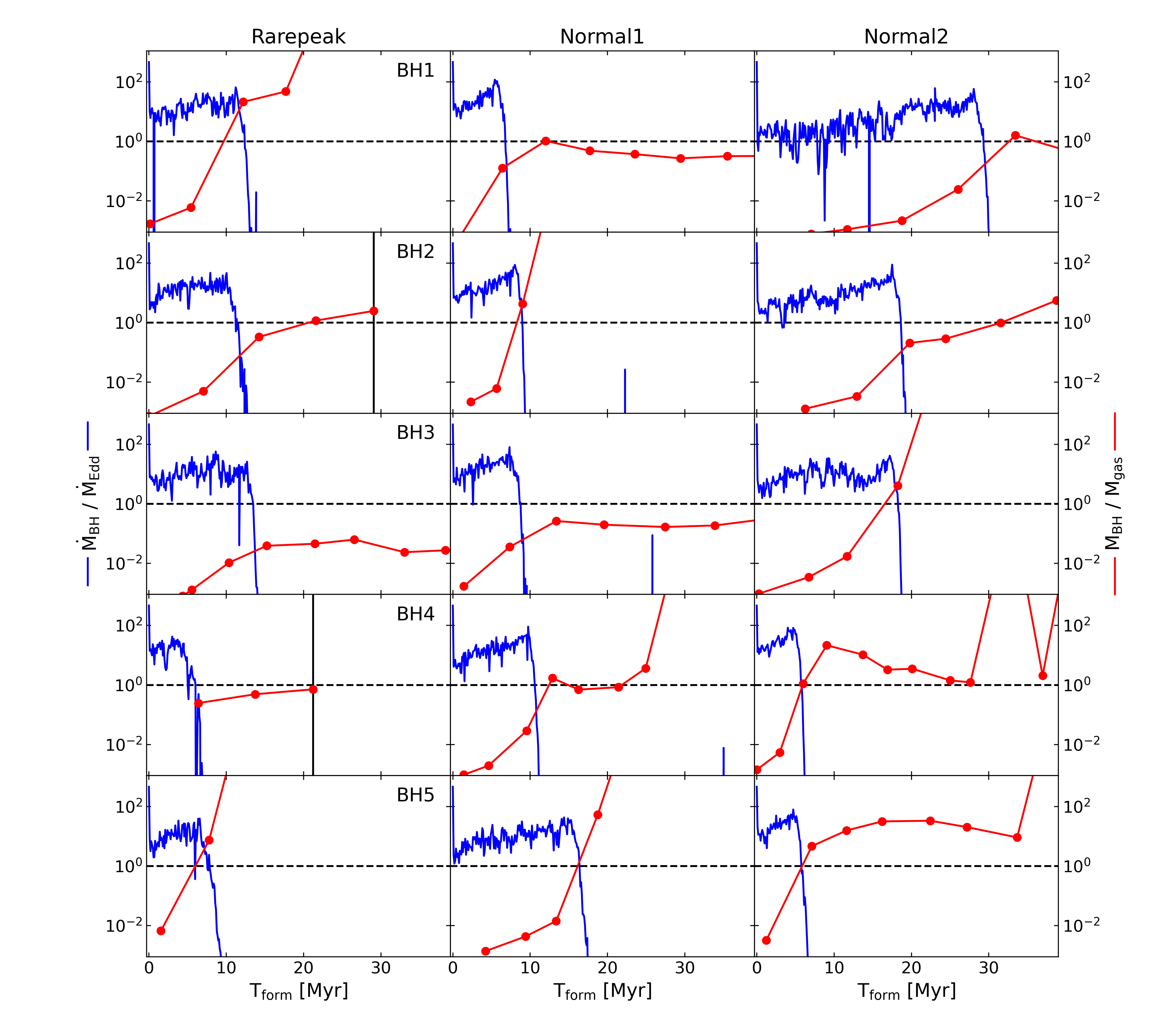}

\caption[]{\label{fig:edd1} For the top 5 most massive MBHs by $z=12.5$ in each simulation, we show the MBH growth rate expressed as the ratio of the accretion rate to the Eddington rate, as a function of time since the formation of the MBH (blue). We also show the ratio of the mass of the MBH to the remaining gas mass within its host halo (red), as identified with the Subfind halo finder. We show the end point of the simulations as a black vertical line where appropriate. \\}
\end{figure*}

\begin{figure*}
  \centering

  \includegraphics[width=\linewidth]{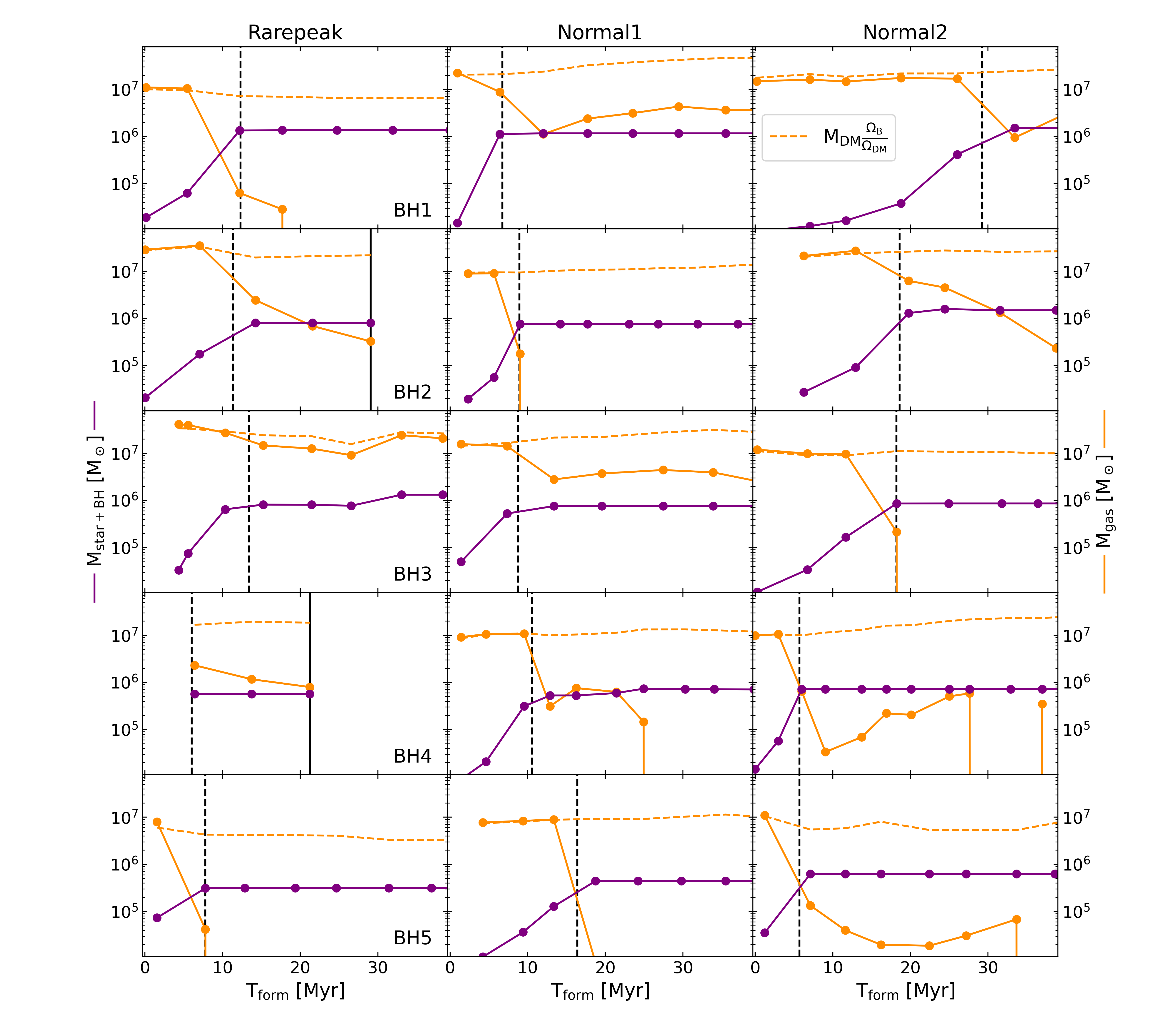}
 
\caption[]{\label{fig:edd2} For the top 5 most massive MBHs by $z=12.5$ in each simulation, we show the combined mass in all stars and BHs within its host halo, as a function of time since the formation of the central MBH (purple). We also show the remaining gas mass within the halo (orange) and compare it to the mass expected from the cosmic ratio of DM to baryons applied to the halo DM mass (orange dashed). Note that this data was taken from snapshots, with a lower output frequency than the black hole data shown in Figure \ref{fig:edd1}. We show the point where the MBH transitions from super- to sub-Eddington growth as a vertical dashed line, and show the end point of the simulations as a black vertical line where appropriate. As MBH accretion shuts off, the combined mass in stars and MBHs does not increase. Therefore, the formation of additional stars and/or MBHs is not responsible for cutting off the primary MBH's accretion supply. \\}
\end{figure*}

\begin{figure*}
  \centering

  \includegraphics[width=0.85\linewidth]{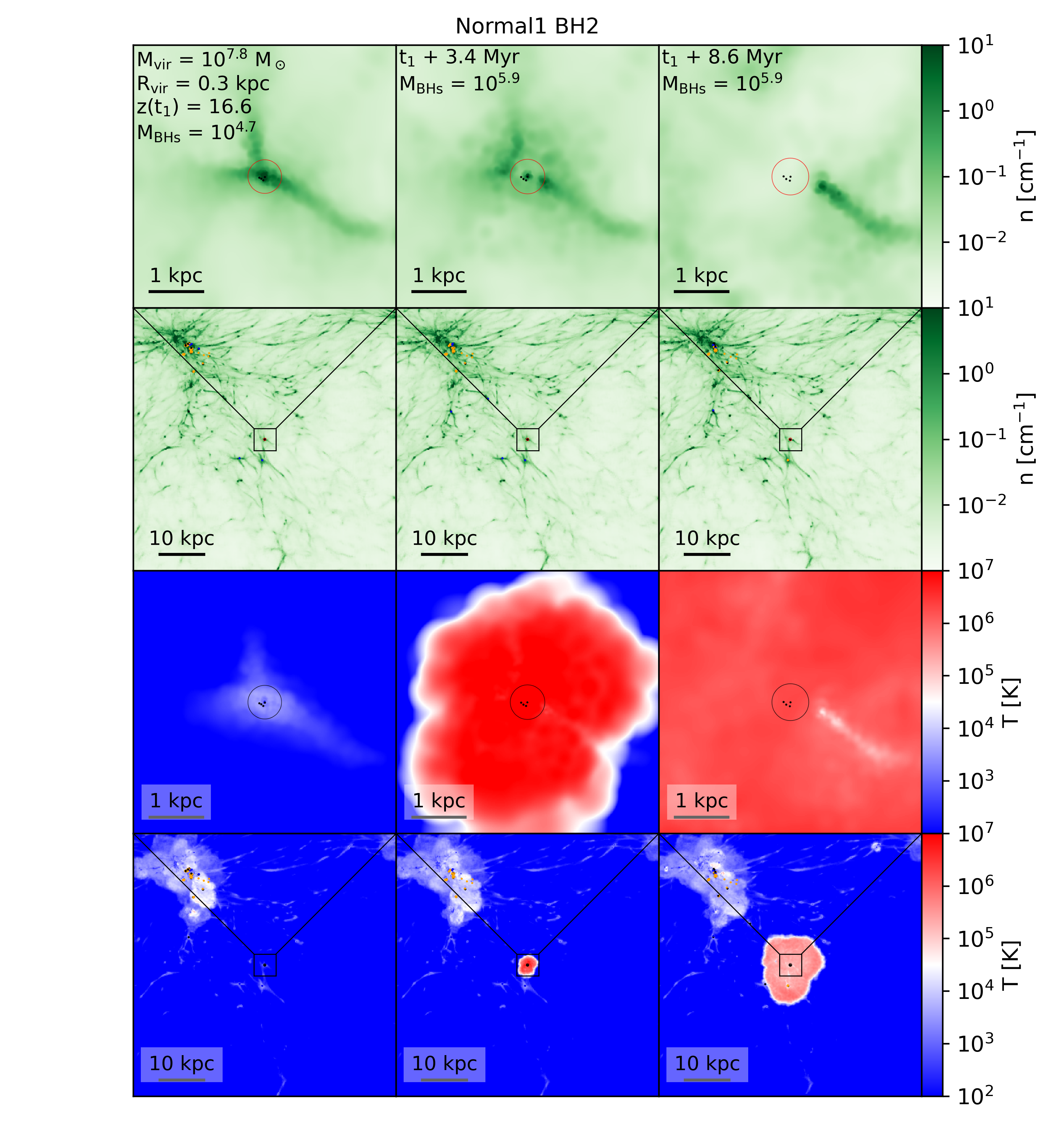}
 
\caption[] 
{\label{fig:N1BH1} Projections of the halo containing N1BH2 before (left), during (center) and after (right) the accretion cut-off. The top 2 panels show density projections, while the bottom 2 panels show temperature projections. Scale lines are given in each panel to indicate the scale of each projection. BHs, PopIII stars and PopII stellar cluster particles are shown as black, blue and orange dots, respectively. We show the half mass radius of the halo as a circle. We include values for the halo mass, radius, redshift and black hole mass. \\ \\
 }
\end{figure*}


\begin{figure*}
  \centering
  \includegraphics[width=0.95\linewidth]{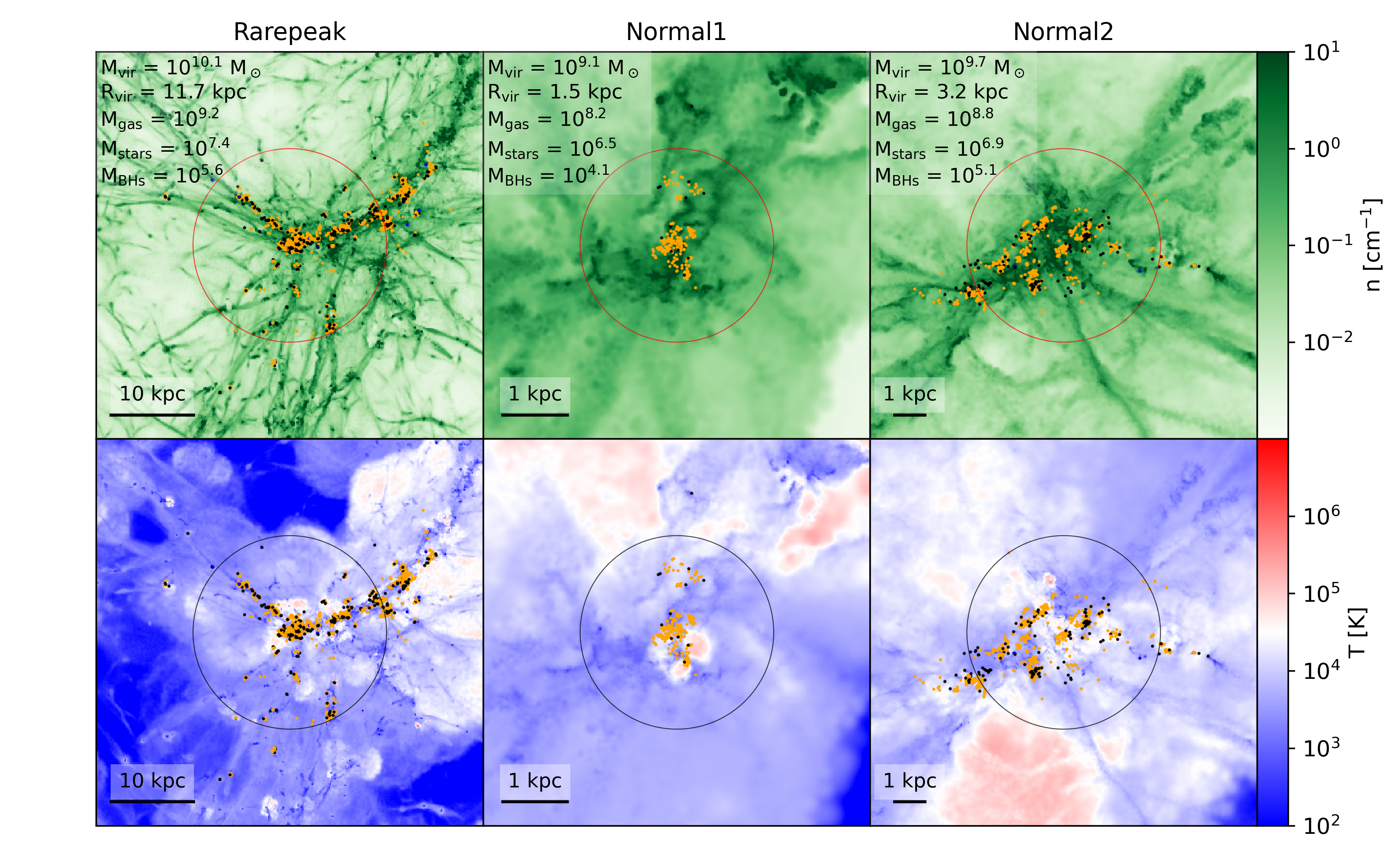}

\caption[]{\label{fig:gasrich} Density (top) and temperature (bottom) projections of the halo with the highest gas mass at $z=12.5$ in each simulation. We match the colorbar to the previous figure to demonstrate the significantly lower temperatures in these halos. BHs, PopIII stars and PopII clusters are shown as black, blue and orange dots, respectively. The halo half mass radius is shown as a circle. We also include values for the halo mass, radius, gas mass, stellar mass and total black hole mass. \\ \\}
\end{figure*}

\begin{figure*}
  \centering
  \hbox{\hspace{1.5cm}
  \includegraphics[width=0.95\linewidth]{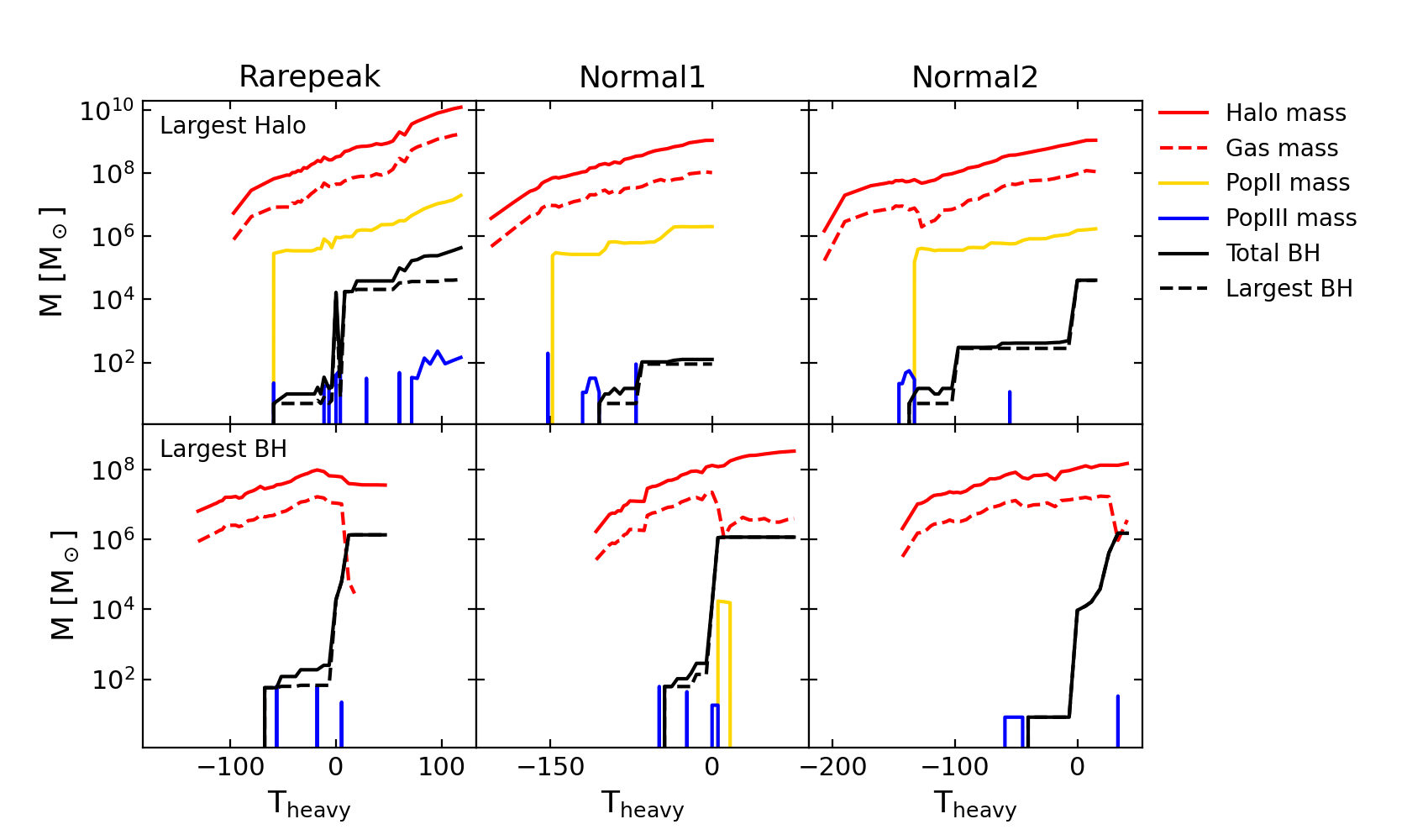}}

\caption[]{\label{fig:SFH} Evolution of the most massive halo in each simultion (top) and the halo hosting the most massive BH in each simulation (bottom). As a function of time since the formation of the first heavy seed BH T$_{\rm heavy}$, we show the halo mass (red), halo gas mass (red dashed), total mass in PopII clusters (orange), total mass in PopIII stars (blue), total mass in BHs (black) and the largest BH mass (black dashed).\\}
\end{figure*}


\begin{figure}
  \centering

  \includegraphics[width=\linewidth]{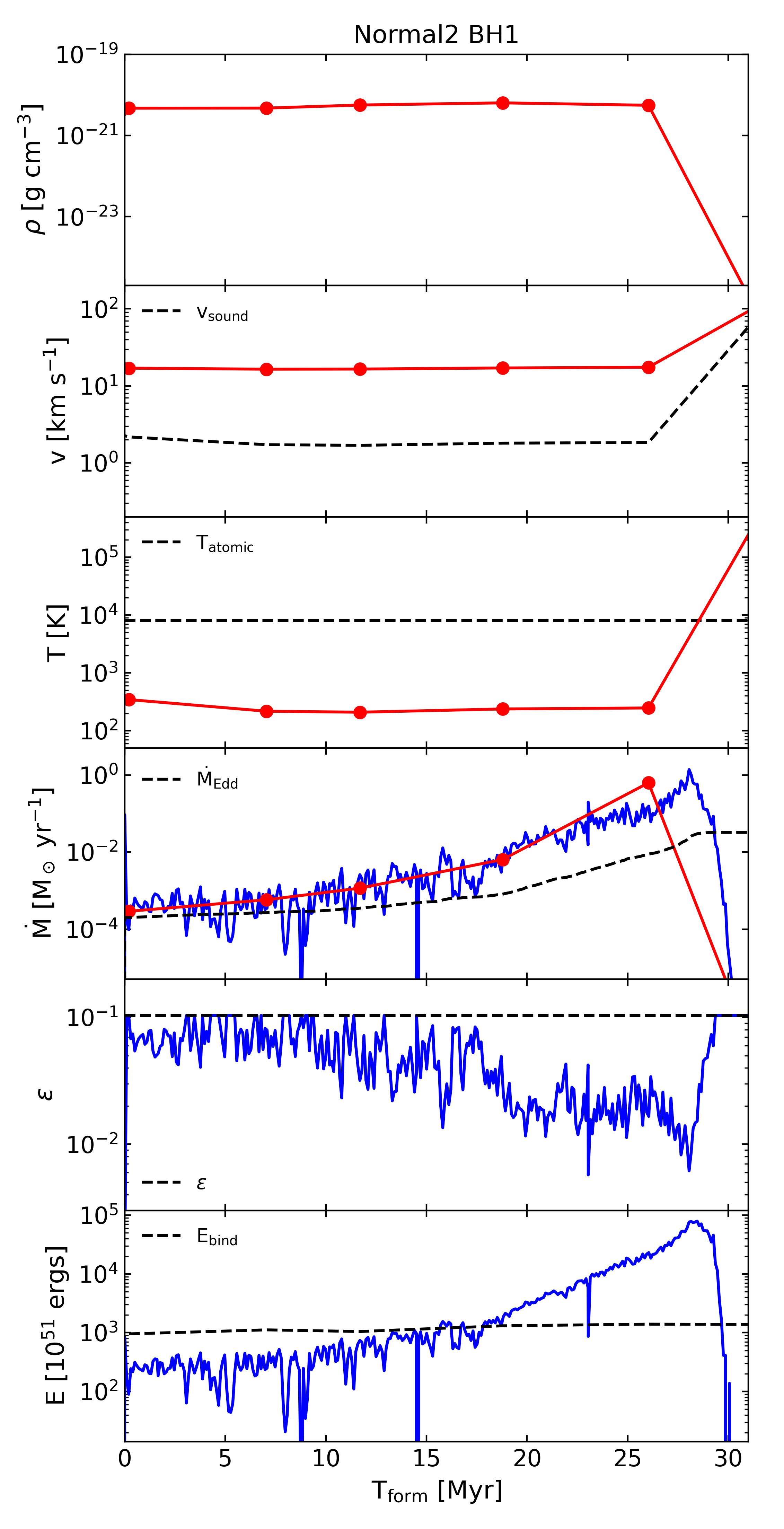}
 
\caption[]{\label{fig:feedback} Accretion and feedback properties of the BH. The top 3 panels show mass weighted average density, relative velocity and temperature of the gas within the accretion sphere surrounding N2BH1. The sound speed is shown as a dashed line. The 4th panel shows the accretion rate onto the BH (blue) and the rate estimated from the properties in the top 3 panels (red). The 5th panel shows the radiative efficiency $\epsilon$ during the super-Eddington burst (blue), compared to the sub-Eddinton value (dashed black). The bottom panel shows the energy injected in each timebin of roughly 10$^5$ yr (blue) compared to the binding energy of the halo (dashed black).}
\end{figure}

\begin{figure*}
  \centering
  \includegraphics[width=0.95\linewidth]{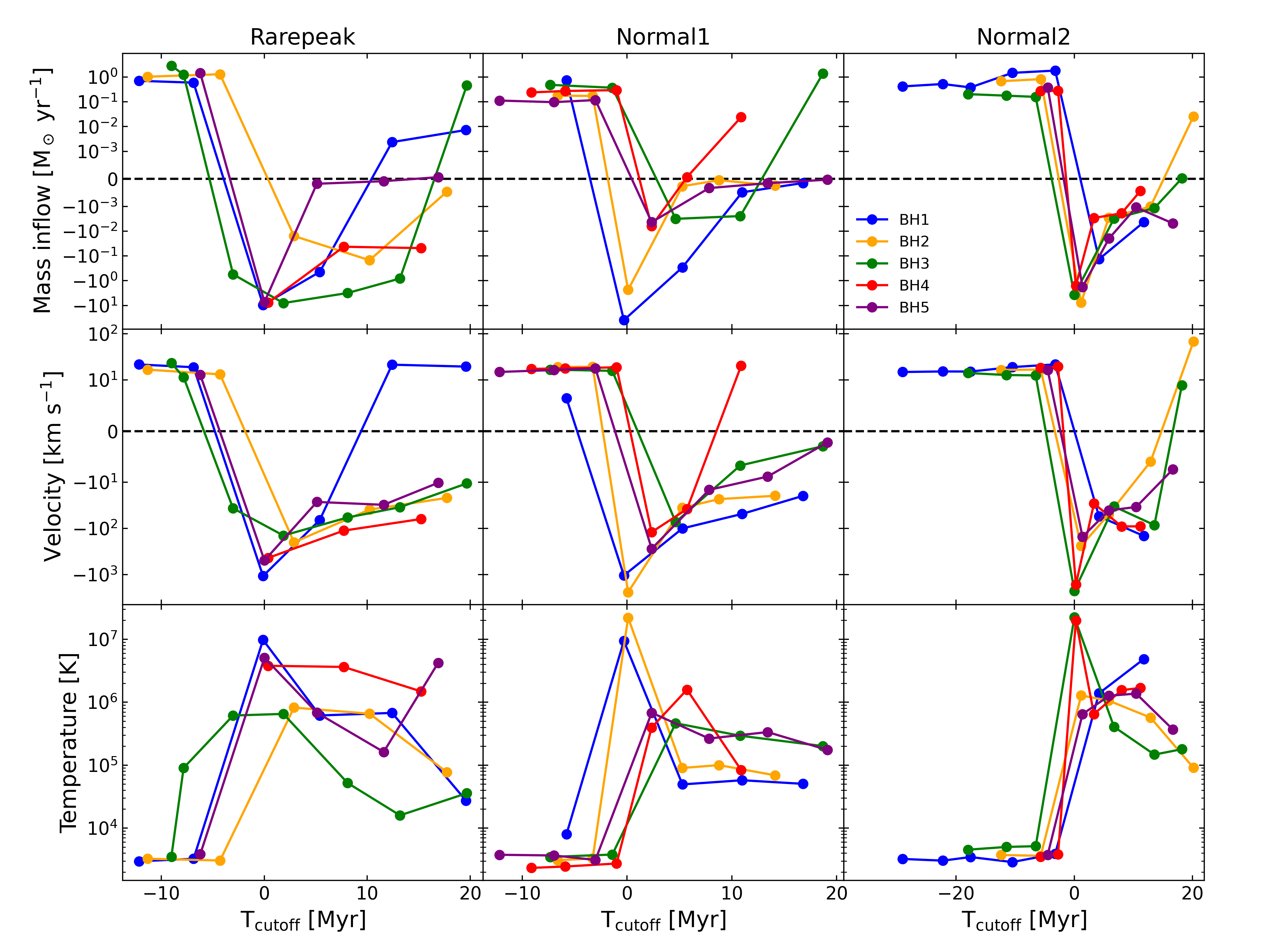}

\caption[]{\label{fig:inflow} The time evolution of halo outflow properties at the time of the MBH accretion cut-off, sampled within a shell extending from the half mass radius R$_{\rm halo}$ out to 1.5R$_{\rm halo}$. Top - radial mass flux (where positive values are net inflow and negative values show net outflowing material), given by the sum of contributions from all cells within the shell. Middle - mass weighted average radial velocities, relative to the central MBH. Bottom - mass weighted average temperatures within the shell. The $x$-axis shows the time before and after the accretion cut-off, T$_{\rm cutoff}$.}
\end{figure*}

\begin{figure*}
  \centering
  \includegraphics[width=0.93\linewidth]{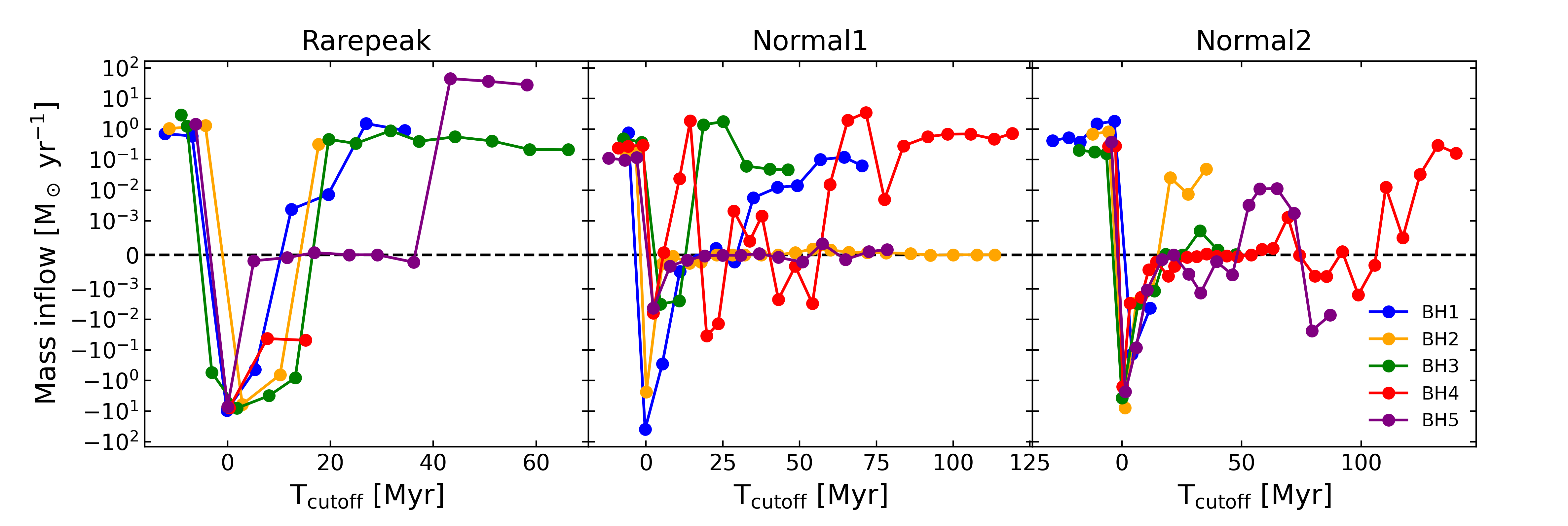}

\caption[]{\label{fig:inflow2} The top panel of Figure \ref{fig:inflow}, but with the $x$-axis extended in time to encompass the time remaining until the end of the simulation. We show gas mass inflow (positive) and outflow (negative) rates, given by the sum of contributions from all cells within a shell extending from R$_{\rm vir}$ out to 1.5R$_{\rm vir}$. The plot shows that many of the MBH's host halos recover from their outflow phase and begin to experience inflow rates comparable to the pre-outflow rates, although none of the MBHs experience a second accretion phase by the end of the simulation.}
\end{figure*}

\section{Results}
\label{sec:results}
\vspace{0.2cm}
In Figure \ref{fig:growth} we show the growth evolution of all heavy seed MBHs in the \rarepeak, \NormalOne and \NormalTwo simulation suites, down to $z = 12.5$. In each case, the most massive black hole reached a mass of just over $10^6$ \msolarc, with most MBHs lying within a factor of a few of their initial seed mass. A significant fraction of the MBHs experienced periods of rapid growth before a sudden accretion cut-off, which is particularly true for the most massive MBHs in each case. \\
\indent As heavy seed MBHs form across a range of masses in the \texttt{SEEDZ} framework, we show explicitly that MBH growth prospects are not set by their initial masses in Figure \ref{fig:initial_mass}, which compares BH masses at $z=12.5$ to their initial seed masses. The data does not indicate that BHs with higher initial masses are more likely to grow efficiently, suggesting that environmental conditions are more relavent to BH growth. \\
\indent We show the halo mass functions (HMFs) in Figure \ref{fig:hmf}, which indicate that our most massive MBHs do not reside in the most massive  halos (10$^9$-10$^{10}$ M$_\odot$), but rather within halos with masses ranging from a few times 10$^7$ M$_\odot$ to a few times 10$^8$ M$_\odot$. Here, we investigate the growth of these MBHs, and defer the examination of the host halo's formation history and their positioning with respect to the greater cosmic web to a follow-up paper. In the following subsections, we investigate the physical mechanism that prevents these MBHs from growing beyond approximately $10^{6}$ \msolarc, by inspecting their environments and host halo properties, before, during and after the accretion cut off.

\subsection{What sets the final mass of the most massive black holes?}
\vspace{0.2cm}

\noindent From a purely numerical perspective, accretion will end when there is a lack of cold, dense gas within the accretion radius of a MBH. There are a number of physical reasons why this may occur: A) MBHs may simply accrete all of the available gas; B) MBH accretion feedback could heat the surrounding gas and severely suppress accretion; C) nearby SNe explosions can evacuate the gas from around a MBH; D) gravitational interactions could eject a MBH from its accretion supply, or E) the dense gas may instead form stars and other MBHs rather than be accreted onto the MBH. In this section, we will disentangle these mechanisms to find out which one (or what combination) drives this process in \texttt{SEEDZ}.

Figure \ref{fig:edd1} shows the MBH growth rate for the top 5 most massive MBHs in each simulation, displayed as the ratio of the accretion rate to the Eddington accretion rate. In each case, super-Eddington accretion lasts for up to a few tens of Myr before experiencing a sharp decrease in growth (similar results have recently been shown by \citealt{Chon2026}). In each case, we identify the MBH's host halo using \texttt{AREPO}'s built-in halo finder, Subfind \citep{Springel2021}. Here groups of at least 20 neighboring DM particles are linked into friends-of-friends parent halos, which are then split into gravitationally bound sub-groups, taking into account the distributions of gas mass and stellar/BH particles. From these bound structures, we take the halo radius $R_{\rm halo}$ to be the half mass radius of the sub-group. The Figure also shows the gas mass within $R_{\rm halo}$ for each of these host halos, which also drops substantially prior to the accretion cut-off in all cases. At the time when the MBHs transitioned from super- to sub-Eddington accretors, the remaining gas mass within each halo has become comparable/much lower than that of the accreting MBH. This raises the question of whether the removed gas mass had been accreted onto the MBH, if the gas had formed into new stars/BHs within the halo, or if the gas was evacuated from each halo by a feedback process. 

To investigate this, Figure \ref{fig:edd2} shows the total mass in all stars and BHs within each of these halos, along with its total remaining gas mass, at the time when accretion was cut off. As the gas mass within each halo decreased, the mass in BHs and stars did not increase to match the deficit, falling short by an order of magnitude. This clearly shows that gas was evacuated from each halo, rather than being accreted or forming new objects.

\subsection{Black hole feedback versus supernovae feedback}
\label{sec:evacuate}
\vspace{0.2cm}
Given that most of the halo's gas was evacuated from the halo at approximately the same time as its central MBH stopped accreting, two mechanisms may have driven this process. Either the thermal accretion feedback injected into the MBH's immediate surroundings heated the gas to such an extent that it expanded out to radii larger than the halo's half mass radius (at very low gas densities), or SN explosions from nearby stars pushed the gas out of the halo via our momentum based SNe prescription. 

To differentiate between these two possibilities, we have identified halos with no stellar formation or SNe activity between the time of MBH formation and the subsequent accretion cut-off. In Figure \ref{fig:N1BH1}  we plot density and temperature projections around N1BH1 (\NormalOne black hole 1), see appendix \ref{fig:N2BH2} and \ref{fig:N2BH4} for more examples from N2BH3 and N2BH4. The panels from left to right show snapshots immediately before, during and after the accretion is cut off. In the absence of SNe explosions, the gas component of the halo was significantly disrupted purely by BH accretion feedback, which heated the gas up to 10$^6$-10$^7$ K on scales larger than the radius of the halo. The region of hot gas grew as the dense gas became more diffuse, spreading out to scales of $\sim$20 kpc before dropping in temperature to 10$^5$ K. The bottom row also encompasses a neighboring halo in the upper left region of the panels, which did experience SNe explosions but lacked a growing MBH. This neighboring halo only reached maximum temperatures of up to $\sim10^5$ K due to SNe feedback alone.

To emphasize that this outflow behavior is caused by MBH feedback rather than from SNe explosions, we also examine halos in our simulations that house large stellar populations. Figure \ref{fig:gasrich} shows projections from the halo with the largest gas mass in each simulation. These halos house BHs which are sub-dominant compared to the stellar mass. In particular, we examine a halo in the \rarepeak simulation which has a mass of $10^{10.1}$ \msolarc, a host gas mass of $10^{9.2}$ \msolarc, a stellar mass of $10^{7.4}$ \msolar and a total BH mass of $10^{5.6}$ \msolarc. In the \NormalOne simulation, the halo has a mass of $10^{9.1}$ \msolar, a host gas mass of $10^{8.2}$ \msolarc, a stellar mass of $10^{6.5}$ and a total BH mass of $10^{4.1}$ \msolarc. Finally, the \NormalTwo halo has a mass of $10^{9.7}$ \msolarc, a host gas mass of $10^{8.8}$ \msolarc, a stellar mass of $10^{6.9}$ \msolar and a total BH mass of $10^{5.1}$ \msolarc. In these halos where the stellar mass is dominant and the BHs are not actively accreting, the galaxies are not quenched and the gas component has not been thermally expanded by SNe feedback compared to the AGN feedback seen in Figure \ref{fig:N1BH1}.

Investigating this further, Figure \ref{fig:SFH} shows the star formation history of the most massive halos in each simulation (top row), and compares them to the halo containing the most massive BH at $z=12.5$ (bottom row), as a function of the time since forming their first heavy seed BH, $T_{\rm heavy}$. It is clear that rapid star formation occurs before ($\sim100$ Myr) the formation of a heavy seed in these massive halos, and continues until the end of the simulation. This prevents BH growth via direct competition for mass, but also from supernovae feedback, which heats the gas and makes accretion less efficient. In contrast, halos containing the most massive BHs experienced very little star formation prior to the formation of a heavy seed BH, allowing for efficient accretion. The resulting AGN feedback from this accretion acts to prevent further star formation in these halos.

\subsection{Galaxy disruption and determining the maximum black hole masses}
\vspace{0.2cm}
It is now clear that MBH accretion feedback causes the sharp accretion cut-off experienced by the MBHs in the \texttt{SEEDZ} simulations. We now post-process the simulation outputs to explain why the MBHs grow so rapidly and why the feedback is so destructive. In Figure \ref{fig:feedback}, we highlight N2BH1, as it has the most snapshot outputs between the formation of the MBH and its accretion cut-off (see Appendix Figure \ref{fig:feedback2} for further examples). The top 3 panels show quantities relevant for calculating the accretion rate using equation \ref{eq:bondi-hoyle-acc}, showing the mass weighted average densities, velocities and temperatures within the accretion sphere surrounding the MBH. Using these values and equation \ref{eq:bondi-hoyle-acc}, we estimate the accretion rate in the 4th panel (red) and compare it to the accretion rate calculated on-the-fly within the simulation (blue), as described in \S \ref{sec:acc}. The combination of high densities, low velocities and temperatures allow for super-Eddington accretion rates, reaching over 10 times the Eddington rate at their peak.

The feedback radiative efficiency $\epsilon$ varies during super-Eddington accretion, given by equation \ref{eq:epsilon}, which is shown in panel 5 of Figure \ref{fig:edd1}. The reduction of $\epsilon$ from its fixed sub-Eddington value allows for weaker feedback during periods of rapid (super-Eddington) growth. This further enables sustained super-Eddington accretion. We use these values of $\epsilon$ with equation \ref{eq:feedback} to estimate the energy injected into the gas surrounding the MBH during the $\sim 10^5$ yr data output intervals, which is plotted in the bottom panel of Figure \ref{fig:feedback}. Note that the estimated energy is likely much larger than the real injected energy due to the injection cut-off described in \S\ref{sec:feedback_method}. As the accretion rate continues to increase, the estimated energy injected during a single $10^5$ yr time interval surpasses the entire binding energy of the halo, which we include as a dashed line for comparison. \\
Halos experience this blowout because the dense gas (responsible for containing the feedback energy and subsequently cooling the region) is consumed through accretion (see top panel of Figure \ref{fig:feedback}). Once this cooling mechanism is lost, coupled to the fact that the energy injected through thermal feedback already exceeds the binding energy of the halos, means that the remaining gas in the halo as blown out, halting accretion for extended periods. \\

The rapid accretion of surrounding dense gas necessary to grow our seed black holes up to $10^6$ \msolar by $z = 12.5$, combined with the binding energy of halos, given by 
\begin{equation}
E_{\rm bind}=\frac{3GM_{\rm vir}^2}{5R_{\rm vir}},
\end{equation}
as well as our non-equilibrium chemistry modelling, effectively sets the maximum mass a MBH can attain before the halo is disrupted. In our simulations, we find that this maximum mass is around $10^6$ M$_\odot$. After this point is reached super-Eddington feedback is sufficient to unbind the gas from its halo (which have masses of at most $10^{10}$ \msolar at this epoch). This halo mass upper limit (as defined by the halo mass function in a $\Lambda$CDM cosmology) is therefore the defining factor in determining the final black hole masses in the \texttt{SEEDZ} simulations. Once the feedback energy released from rapidly accreting MBHs can not be contained by the gravitational energy of their host halos, the gas is fully evacuated and accretion stops. We therefore predict that galaxies at the epoch of $z \sim 10$ may show signs of significant quenching, unless the accretion feedback can be efficiently damped near to the MBH.

\subsection{Halo outflows and long-term recovery}
\vspace{0.2cm}
For a broader investigation into the outflow behavior discussed in \S\ref{sec:evacuate}, we plot inflow/outflow properties around each of the 5 most massive MBH's host halos in Figure \ref{fig:inflow}. We take averaged quantities within a shell spanning from the halo half mass radius $R_{\rm halo}$, out to 1.5$R_{\rm halo}$, at the time of the accretion cutoff, T$_{\rm cutoff}$. We show the mass inflow/outflow rate in the top panel of Figure \ref{fig:inflow} - in all cases, initial inflow rates were between $0.1-1$ M$_\odot$ yr$^{-1}$. At the time of the accretion cut-off, this positive inflow switched to a negative outflow, with highs of $10^{-1}$ M$_\odot$ yr$^{-1}$, before slowing down towards a near stationary flow (0 M$_\odot$ yr$^{-1}$) on timescales of 5-10 Myr. This indicates that feedback from all of the considered MBHs thermally expanded the gas component outside of their host halo's radii. To make predictions for possible observations of such outflows surrounding growing MBHs, we plot the outflow velocities and temperatures in the middle and bottom panels. At the point where accretion ends, gas outflows at 10$^3$ km s$^{-1}$, generally slowing to 10 km s$^{-1}$ on timescales of 10 Myr. These outflows reach temperatures of 10$^7$ K at their peak, cooling to 10$^5$ K in the following 10-20 Myr.\\
\indent Despite exhibiting prominent outflows, many of these halos do eventually recover and reach inflow rates comparable to the pre-outflow values. Figure \ref{fig:inflow2} shows the mass inflow rates over a longer timeframe, running from the MBH's creation to the end of the simulation at $z=12.5$. Some of these systems recover inflow rates of 0.1 M$_\odot$ yr$^{-1}$ on short timescales of 20-40 Myr (\rarepeak BH1,3,5 and \NormalOne BH1,3,4), while others recover after roughly 100 Myr (\NormalOne BH4, \NormalTwo BH4). However, for many of these MBHs, there is not enough time between the outflow period and the end of the simulation to begin inflowing. For example \NormalOne BH2 and BH5 remain at net zero mass inflow for almost 100 Myr. Even in systems that do regain their high inflow rates, the MBH does not experience another growth phase by the end of the simulation. However, MBH accretion is likely to re-occur if given more time. As our simulations continue to evolve, we will return to this point in a follow-up work.

\section{Discussion \& conclusions}
\vspace{0.2cm}
In this paper, we have analysed the growth and feedback properties of the most massive BHs in the \texttt{SEEDZ} simulations, yielding a number of important conclusions, listed  below:
\label{sec:conclusions}
   \begin{itemize}
      \item In a cosmological setting, heavy seed BHs with initial masses between $5 \times 10^3$ and $10^5$ \msolar are capable of growing into 10$^6$ M$_\odot$ SMBHs by $z=12.5$
      \item The most massive MBHs gain their mass during short periods of super-Eddington accretion lasting between 5-30 Myr.
      \item The most active and massive MBHs in our suite inject significant thermal energy into their surroundings to unbind the gas from its host halo. The rapid accretion and resulting strong feedback effectively quenches the host galaxy. This sets a maximum MBH mass at this epoch (determined by the binding energy and hence mass of the host halo). Our simulations predict maximum MBH masses of approximately $10^6$ \msolar at $z = 12.5$.
      \item Either MBH feedback in the early Universe is much mess efficient than our model, or the MBHs we observe with JWST may have evacuated their host halo's gas reservoir during their formation, and later built up a stellar component following replenishment of their gas reservoir from the cosmic web or through halo mergers.
      
   \end{itemize}
\vspace{0.1cm}
The goal of this paper was to examine the growth and limitations of MBHs in the \texttt{SEEDZ} simulation suite. Our main finding is that MBHs are unlikely to exceed approximately $10^6$ \msolar through accretion alone by $z = 12.5$. The associated feedback injected into the inter-stellar medium due to rapid accretion unbinds the gas from the host halo. In principle, this tells us that any future observations of MBHs with masses significantly larger than $10^6$ \msolar before $z=12.5$ would have major implications for our models of the early Universe. Of course, this assumes that our model for MBH accretion and feedback captures the key physical processes and is numerically converged.

The feedback model implemented in the \texttt{SEEDZ} simulations injects thermal feedback isotropically into the surrounding medium during accretion episodes. It is entirely possible that our feedback model and parameters are too strong. However, the feedback prescription is already relatively weak, employing thermal feedback with a gas coupling factor of 0.05, and further reducing the radiative efficiency $\epsilon$ during super-Eddington phases, as described in \S \ref{sec:feedback_method}. In a forthcoming study, we will explore the parameter space of these feedback parameters. \\
\indent An additional consideration is that merger dominated growth may help to alleviate the negative consequences of accretion driven feedback. This current set of simulations is too coarse to fully evaluate the impact of the mergers and to what extent they can dominate the growth the MBHs at high redshift. Subsequent \texttt{SEEDZ} simulations suites at higher resolution will go some way to tackling this problem. Allied to this, our results are broadly consistent with the \texttt{BRAHMA} simulation suite \citep{Bhowmick2026}, where AGN feedback was also shown to suppress early black hole growth.

\subsection{Observational Guidance from Little Red Dots}
\vspace{0.2cm}
\noindent Recent observations of so-called Little-Red-Dot (LRD) \citep[][]{Matthee2024} systems offer potential observational guidance to our theoretical models. If LRDs are systems with dense gas surrounding a rapidly accreting MBH at the center \citep[e.g.][]{Inayoshi2025a,Inayoshi2025, Naidu2025, DeGraaff2025, DeGraff2025a, Rusakov2025, Rusakov2026, Sun2026} as they appear to be, then our feedback modeling may require some adjustment. The observed systems hosting LRDs have total masses not significantly beyond our largest halo masses, in the region of approximately $10^{11}$ \msolar (our most massive halos will reach masses in excess of this by z = 10). Our modeling suggests that such halos should show significant signs of galaxy quenching. However, observational evidence is still not fully converged. For example, \cite{Sneppen2026} suggest that dense gas surrounding growing MBHs is able to effectively contain feedback energy. On the other hand, recently there have been several reports of AGN outflows from LRDs and high-z galaxies \citep[e.g.][]{Korber2026, Wang2025}. \\
\indent The increased resolution in our upcoming suite of follow-up simulations may result in LRD-like systems, as the increased resolution will allow for higher densities surrounding growing MBHs and therefore a possibly different feedback result. Nonetheless, early observations from LRDs (assuming they host small but actively accreting MBHs) suggests that feedback in these early galaxies does not always disrupt the host halo, despite the necessarily high accretion rates. Hence, higher resolution models coupled with further observational guidance will be required to understand MBHs in the early Universe. 

\subsection{Comparison to other high redshift models}
\vspace{0.2cm}
\noindent A detailed comparison between the \texttt{SEEDZ} framework and other cosmological simulations was presented in \cite{Prole2026}, here we focus briefly on the most relevant recent works for interpreting our results and conclusions.  \\
\indent The \texttt{BRAHMA} \citep{Bhowmick2024, Bhowmick2024a, Bhowmick2025, Bhowmick2025b, Bhowmick2026} and \texttt{MELI$\odot$RA} \citep{Cenci2025} simulation suites represent the closest points of comparison, as both include physically motivated black hole seeding prescriptions and are designed to probe high-redshift black hole growth. However, key differences in subgrid modeling remain. In particular, both \texttt{BRAHMA} and \texttt{MELI$\odot$RA} employ effective equation-of-state (eEOS) treatments for the ISM, which do not explicitly resolve the cold gas phase, and adopt star formation, MBH accretion and feedback prescriptions more similar to those used in large-volume simulations. \\
\indent In contrast, \texttt{SEEDZ} follows non-equilibrium chemistry and resolves the multiphase ISM, allowing dense, cold gas to form and persist. This has important implications for black hole growth. In models employing an eEOS approach feedback leaves gas heated and dispersed, which can suppress sustained accretion and favor growth via black hole mergers. By comparison, \texttt{SEEDZ} allows prolonged inflow of cold, dense gas, leading to accretion-dominated growth at early times. However, this conclusion should be tempered in light of the suppression of mergers in the current \texttt{SEEDZ} suite due to resolution constraints. \\
\indent  One of the key differences between eEOS modeling and the use of a detailed chemistry model is the emergence of a rapid initial accretion phase in \texttt{SEEDZ}, which eEOS models can not produce because the cold gas phase is not resolved. Following this initial stage of rapid accretion (which lasts only a few Myr) the masses of the MBHs in \texttt{SEEDZ} are between approximately $10^4$ M$_\odot$ and $10^6$ M$_\odot$, with all of the growth coming through accretion. It therefore follows that the initial seeds in \texttt{BRAHMA} and \texttt{MELI$\odot$RA} may represent the post-accretion, post-AGN-feedback phase of the MBHs in \texttt{SEEDZ}. It is therefore possible that in this later phase of the MBH's cycle, mergers may become the dominant growth mechanism (which is observed in \texttt{BRAHMA}). The first generation of \texttt{SEEDZ} simulations presented here do not resolve this merger phase sufficiently well to explore this regime, but this will be investigated in a future, higher resolution suite (Mehta et al. in prep). A possible ideal endpoint is that the explicit chemistry modelling during the accretion phase in the \texttt{SEEDZ} suite is used to parameterize the early growth phases of MBHs, which can then be used as the starting point of MBH seeding in larger scale, more computationally affordable simulations, extendable to lower redshifts.

\section*{Acknowledgements}
\noindent JR \& JB acknowledges support from the Royal Society and Research Ireland under grant number 
 URF\textbackslash R1\textbackslash 191132. LP, DM \& JR acknowledge support from the Research Ireland Laureate programme under grant number IRCLA/2022/1165. JHW acknowledges support from NSF grants AST-2108020 and AST-2510197 and NASA grant 80NSSC21K1053. RSB acknowledges support from a UKRI Future Leaders Fellowship MR/Y015517/1.
 \ \
The simulations were performed on the Luxembourg national supercomputer MeluXina and the Czech Republic EuroHPC machine Karolina.
The authors gratefully acknowledge the LuxProvide teams for their expert support.
\ \ 
The authors wish to acknowledge the Irish Centre for High-End Computing (ICHEC) for the provision of computational facilities and support.
\ \
The authors acknowledge the EuroHPC Joint Undertaking for awarding this project access to the EuroHPC supercomputer Karolina, hosted by IT4Innovations through a EuroHPC Regular Access call (EHPC-REG-2023R03-103) and to the LuxProvide supercomputer Meluxina  through a EuroHPC Regular Access call (EHPC-REG-2025R01-008).
\ \ 
S.K. acknowledges support from the Royal Society under grant number URF\textbackslash R1\textbackslash 251867 and the Royal Commission for the Exhibition of 1851.
\ \
M.A.B. is supported by a UKRI Stephen Hawking Fellowship (EP/X04257X/1)
\ \ 
RSK acknowledges financial support from the ERC via Synergy Grant ``ECOGAL'' (project ID 855130) and from the German Excellence Strategy via the Heidelberg Cluster ``STRUCTURES'' (EXC 2181 - 390900948). In addition RSK is grateful for funding from the German Ministry for Economic Affairs and Climate Action in project ``MAINN'' (funding ID 50OO2206), and from DFG and ANR for project ``STARCLUSTERS'' (funding ID KL 1358/22-1).

\bibliographystyle{aasjournal}
\bibliography{references_seedz}

\appendix
\section{More examples of BH feedback}
\label{sec:appendix}
\counterwithin{figure}{section}
Here we include more examples of plots presented within the main body of the paper, to strengthen the arguments made throughout the text. 

\begin{figure*}
  \centering

  \includegraphics[width=\linewidth]{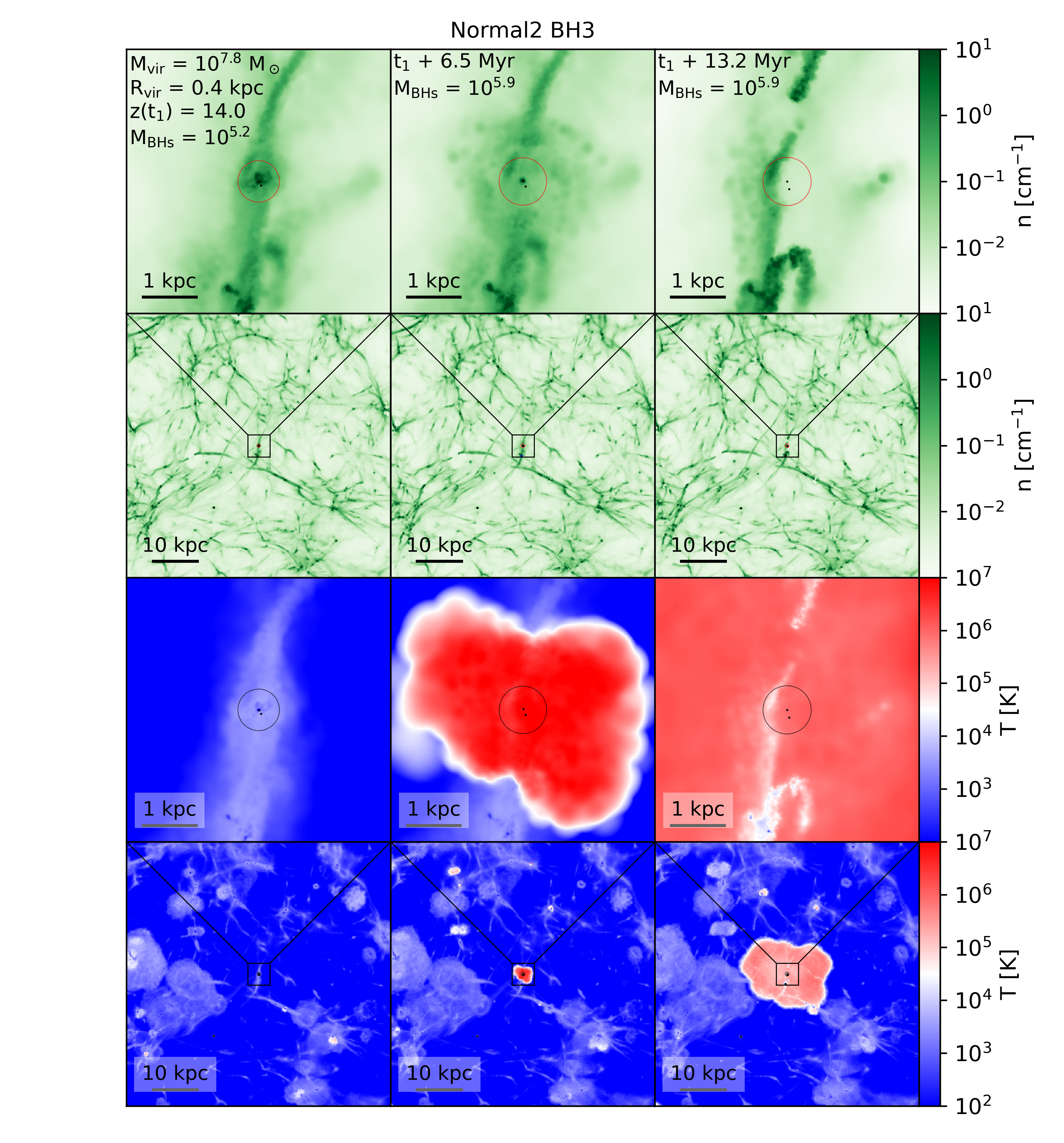}
 
\caption[] {\label{fig:N2BH2}Projections of the halo containing N2BH3 before (left), during (center) and after (right) the accretion cut-off. The top 2 panels show density projections, while the bottom 2 panels show tempertaure projections. In each case, the upper panel shows a region of 5 kpc, while the lower panel shows a zoom-out of 60 kpc. BHs, PopIII stars and PopII stellar cluster particles are shown as black, blue and orange dots, respectively. We show the half mass radius of the halo as a circle. We include values for the halo mass, radius, redshift and black hole mass.
 }
\end{figure*}

\begin{figure*}
  \centering

  \includegraphics[width=\linewidth]{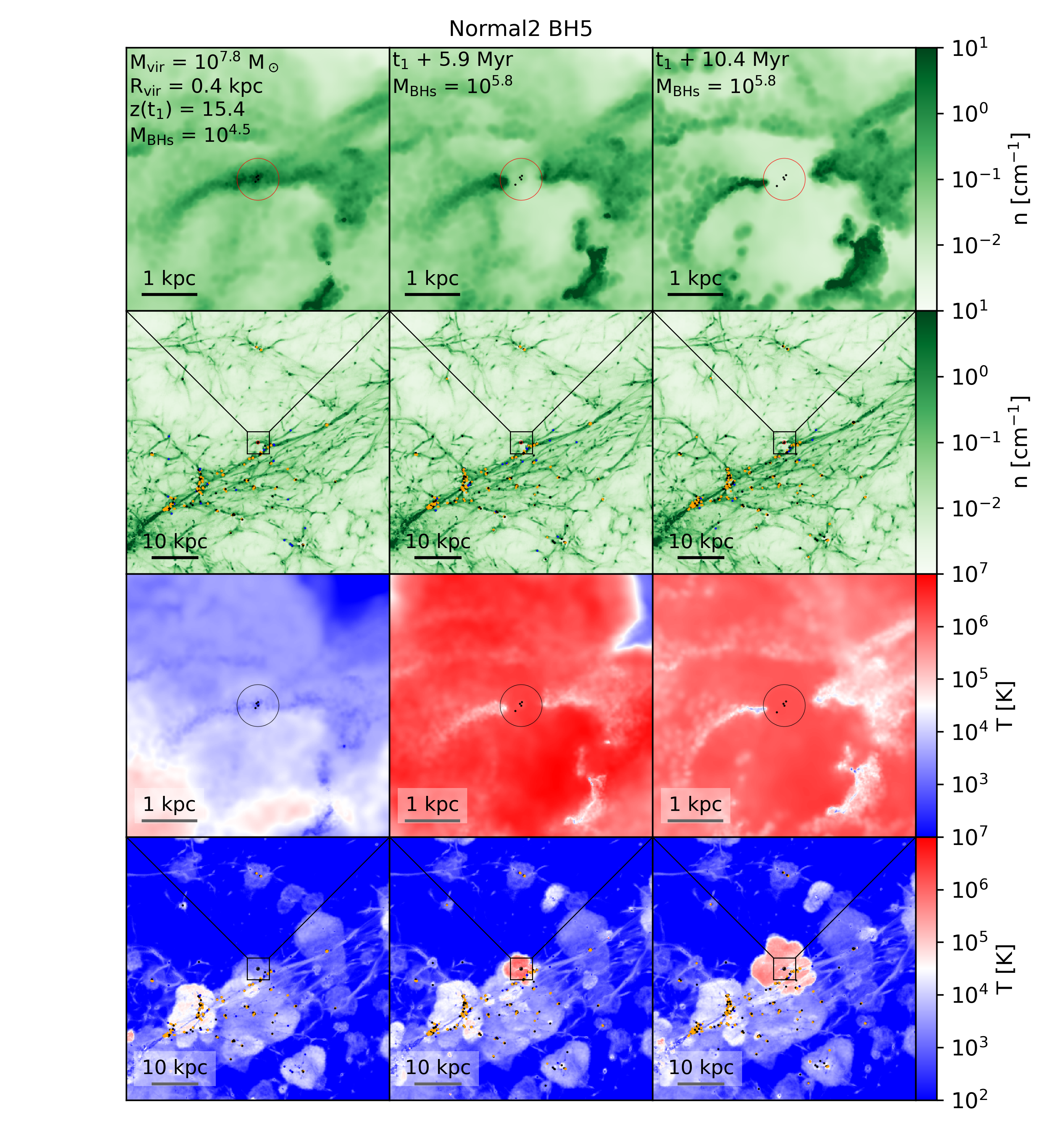}
 
\caption[] {\label{fig:N2BH4}Projections of the halo containing N2BH5 before (left), during (center) and after (right) the accretion cut-off. The top 2 panels show density projections, while the bottom 2 panels show tempertaure projections. In each case, the upper panel shows a region of 5 kpc, while the lower panel shows a zoom-out of 60 kpc. BHs, PopIII stars and PopII stellar cluster particles are shown as black, blue and orange dots, respectively. We show the half mass radius of the halo as a circle. We include values for the halo mass, radius, redshift and black hole mass.
 }
\end{figure*}

Firstly, we show similar density and temperature projections to Figure \ref{fig:N1BH1} in Figures \ref{fig:N2BH2} and \ref{fig:N2BH4}, which correspond to the host halos of N2BH2 and N2BH4, respectively. These 3 systems were chosen because they contain the only MBHs from the sample of 15 largest BHs inspected to experience no SNe explosions between the time of their creation and the eventual accretion cut-off. Therefore, the feedback effects can only be attributed to BH accretion feedback. In all cases, feedback from the MBH heats the gas to 10$^{7}$ K, expanding it beyond the radius of the halo, out to scales of 10 kpc.

\begin{figure}%
    \centering
    \includegraphics[width=0.49\linewidth]{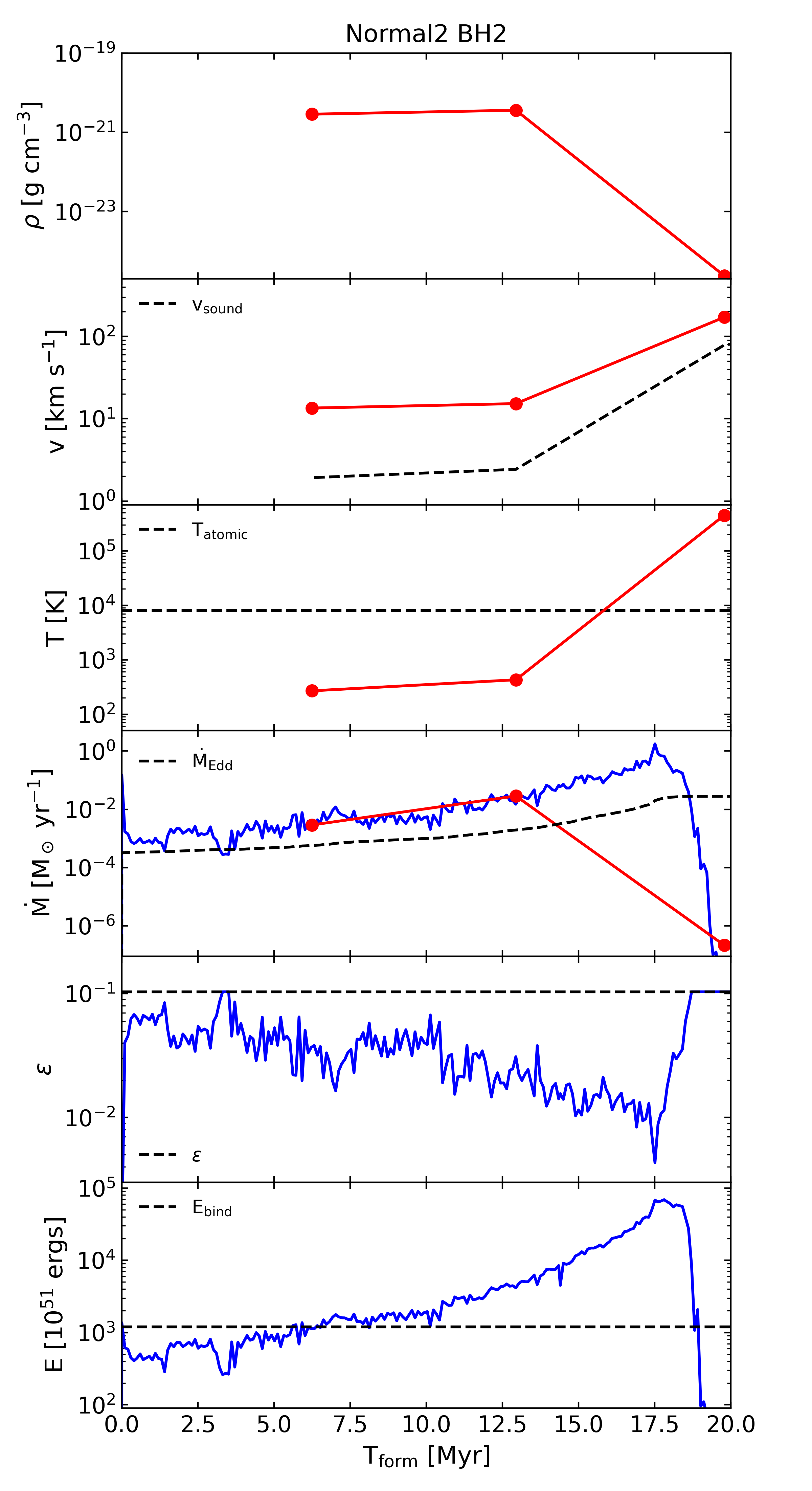} %
    \includegraphics[width=0.49\linewidth]{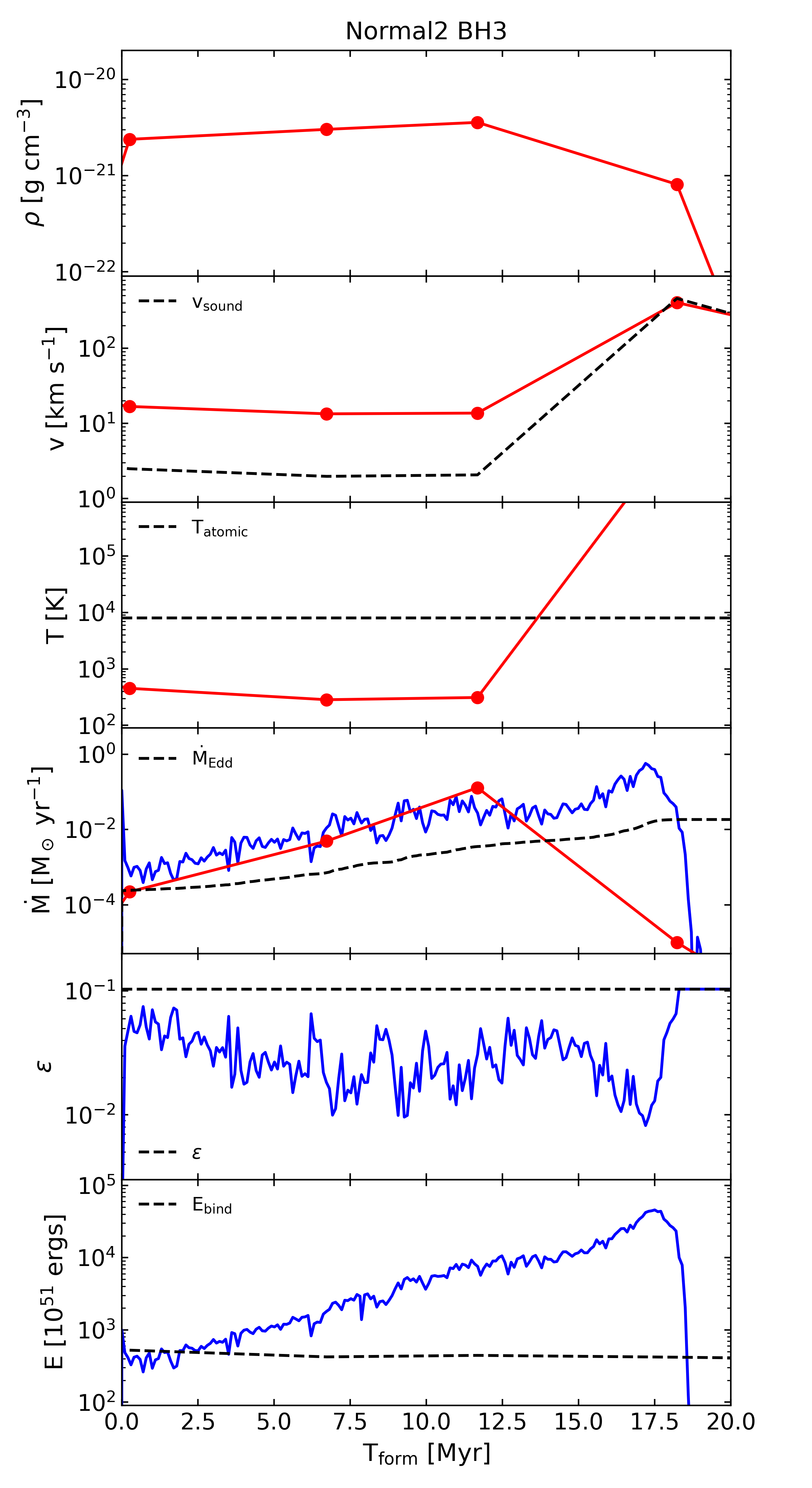} %
    \caption{Accretion and feedback properties of N2BH2 and N2BH3. The top 3 panels show mass weighted average density, relative velocity and temperature of the gas within the accretion sphere surrounding the BH. The sound speed is shown as a dashed line. The 4th panel shows the accretion rate onto the BH (blue) and the rate estimated from the properties in the top 3 panels (red). The 5th panel show the feedback parameter $\epsilon$ during the super-Eddington burst (blue), compared to the sub-Eddinton value (dashed black). The bottom panel shows the energy injected in each timebin of roughly 10$^5$ yr (blue) compared to the binding energy of the halo (dashed black).}%
    \label{fig:feedback2}%
\end{figure}

In the main text, we presented the gas properties within the accretion sphere of N2BH1 and the estimated feedback energies in Figure \ref{fig:feedback}, because it had the most snapshot outputs between the BH's creation and its accretion cut-off. Here we present similar plots for N2BH2 and N2BH3 in Figure \ref{fig:feedback2}, as these have the next highest number of snapshots during the accretion period. In all 3 cases, the accretion cut-offs occur approximately 15 Myr after the injected energy per 10$^5$ Myr timestep exceeds the binding energy of the halo. 


\end{document}